\documentclass{article}

\usepackage{amsmath}
\usepackage{amsfonts}
\usepackage{epsfig}
\usepackage{graphicx}

\newtheorem{theorem}{Theorem}

\newtheorem{corollary}{Corollary}

\newtheorem{lemma}{Lemma}
\newtheorem{example}{Example}

\newtheorem{remark}{Remark}

\def\min{\mathop{\rm min}}

\def\Iden{\mbox{$\bf 1\ $}}

\def\Ze{\mbox{$\bf 0\ $}}
\def\n{\noindent}
\def\no{\noindent}

\def\h {\mathfrak{h}}

\def\a {\mathfrak{a}}

\def\p {\mathfrak{p}}

\def\g {\mathfrak{g}}
\def\k {\mathfrak{k}}
\def\f {\mathfrak{f}}
\def\m {\mathfrak{m}}
\def\z  {\mathfrak{z}}

\def\m  {\mathfrak{m}}

\def\W {\mathcal {W}}
\def\P {\mathcal {P}}
\def\c {\mathfrak{c}}
\def\C {\mathcal{C}}
\def\B {\mathcal{B}}
\def\I {\mathcal {I}}

\begin{document}

\baselineskip .7cm

\author{ Navin Khaneja \thanks{To whom correspondence may be addressed. Email:navinkhaneja@gmail.com} \thanks{Department of Electrical Engineering, IIT Bombay - 400076, India.}}

\vskip 4em

\title{\bf Time optimal control in coupled spin systems: a second order analysis}

\maketitle

\vskip 3cm

\begin{center} {\bf Abstract} \end{center}

In this paper, we study some control problems that derive from time optimal control of coupled spin dynamics in NMR spectroscopy and 
quantum information and computation. Time optimal control helps to minimize relaxation losses. The ability to synthesize, local unitaries, much 
more rapidly than evolution of couplings, gives a natural time scale separation in these problems. The generators of evolution, $\g$, are decomposed 
into fast generators $\k$ (local Hamiltonians) and slow generators $\p$ (couplings) as a Cartan decomposition $\g = \p \oplus \k$. Using this  
decomposition, we exploit some convexity ideas to completely characterize the reachable set and 
time optimal control for these problems. In this paper, we carry out a second order analysis of time optimality.

\vskip 3cm







\section{Introduction}

A rich class of model control problems arise, when one considers dynamics of two coupled spin
$\frac{1}{2}$. The dynamics of two coupled spins, forms the basis
for the field of quantum information processing and
computing \cite{nc} and is fundamental in multidimensional NMR spectroscopy \cite{Ernst}, \cite{Palmer}.
Numerous experiments in NMR spectroscopy, involve synthesizing unitary transformations \cite{timeopt, cartan, Alessandro} that
require interaction between the spins (evolution of the coupling Hamiltonian). These
experiments involve transferring, coherence and polarization from one spin to another and involve
evolution of interaction
Hamiltonians \cite{Ernst}. Similarly,
many protocols in quantum communication and information processing involve synthesizing
entangled states starting from the
separable states \cite{nc, kraus, bennett}. This again requires evolution of interaction Hamiltonians between the
qubits.

A typical feature of many of these problems is that evolution of interaction
Hamiltonians takes significantly longer than the time required to generate local
unitary transformations (unitary transformations that effect individual spins only). 
In NMR spectroscopy \cite{Ernst, Palmer}, local unitary transformations on spins are obtained by application
of rf-pulses, whose strength may be orders of magnitude larger than  the
couplings between the spins. Given the Schr\'oedinger equation for unitary evolution
\begin{equation}
\label{eq:sfcoupling}
\dot{U} = -i [H_c + \sum_{j=1}^n u_j H_j] U , \ \ U(0) = I,
\end{equation} where
$H_c$ represents a coupling Hamiltonian, and $u_j$ are controls that can be switched on and off. What is
the minimum time required to synthesize any
unitary transformation in the coupled spin system, when the control generators
$H_j$ are local Hamiltonians and are much stronger than the coupling between
the spins ($u_j$ can be made large). Design of time optimal rf-pulse sequences is an important research subject
in NMR spectroscopy and quantum information processing \cite{timeopt}-\cite{hai3spin}, as minimizing the time to execute
quantum operations can reduce relaxation losses, which are always present in an
open quantum system \cite{Redfield, Lind}.
The
present problem has a special mathematical structure that helps to characterize all
the time optimal trajectories \cite{timeopt}. The special mathematical structure manifested
in the coupled two spin system, motivates a broader study of control systems
with the same properties.

The Hamiltonian of a spin $\frac{1}{2}$ can be written in terms of the generators of
rotations on a two dimensional space and these are the Pauli matrices
$-i \sigma_x, -i \sigma_y , -i \sigma_z$, where,
\begin{equation}
\sigma_z = \frac{1}{2} \left [ \begin{array}{cc} 1 & 0 \\ 0 & -1 \end{array} \right ]; \ \
\sigma_y = \frac{1}{2} \left [ \begin{array}{cc} 0 & -i \\ i & 0 \end{array} \right ]; \ \
\sigma_x = \frac{1}{2} \left [ \begin{array}{cc} 0 & 1 \\ 1 & 0 \end{array} \right ]. \ \
\end{equation}Note
\begin{equation}
\label{eq:paulicommute}
[\sigma_x, \sigma_y ] = i \sigma_z, \ \  [\sigma_y, \sigma_z ] = i \sigma_x, \ \ [\sigma_z, \sigma_x ] = i \sigma_y,
\end{equation}where $[A, B] = AB-BA$ is the matrix commutator and
\begin{equation}
\label{eq:pauliproduct}
\sigma_x^2 = \sigma_y^2 = \sigma_z^2 = \frac{\Iden}{4},
\end{equation}

The Hamiltonian for a system of two coupled spins takes the general form

\begin{equation}
\label{eq:naturalH}
H_0 = \sum a_{\alpha} \sigma_{\alpha}\otimes \Iden + \sum b_{\beta} \Iden \otimes  \sigma_{\beta}
+ \sum J_{\alpha \beta} \ \sigma_{\alpha} \otimes \sigma_{\beta},
\end{equation} where $\alpha, \beta \in \{x, y, z \}$. The Hamiltonians $\sigma_{\alpha}\otimes \Iden $ and
$\Iden \otimes \sigma_{\beta}$ are termed local Hamiltonians and operate on one of the spins. The Hamiltonian
\begin{equation}
\label{eq:couplingH}
H_c = \sum J_{\alpha \beta} \ \sigma_{\alpha} \otimes \sigma_{\beta},
\end{equation}is the coupling or interaction Hamiltonian and
operates on both the spins.

The following notation is therefore common place in the NMR literature.
\begin{equation}
I_{\alpha} = \sigma_{\alpha} \otimes \Iden \ \ ;\ \ S_{\beta} = \Iden \otimes \sigma_{\beta}.
\end{equation} The operators $I_{\alpha}$ and $S_{\beta}$ commute and therefore
$\exp( -i \sum_{\alpha} a_\alpha I_\alpha + \sum_{\beta} b_{\beta} S_{\beta} )=$
\begin{equation}
\exp( -i \sum_{\alpha} a_\alpha I_\alpha)\exp(-i \sum_{\beta} b_{\beta} S_{\beta} )= (\exp( -i \sum_{\alpha} a_\alpha \sigma_\alpha)\otimes \Iden)(\Iden \otimes \exp(-i \sum_{\beta} b_{\beta} \sigma_{\beta} ),
\end{equation}The unitary transformations of the kind
$$  \exp( - i \sum_{\alpha} a_\alpha \sigma_\alpha )\otimes \exp( -i \sum_{\beta} b_\beta \sigma_\beta ), $$
obtained by evolution of the local Hamiltonians are called local unitary transformations.

The coupling Hamiltonian can be written as
\begin{equation}
H_c = \sum J_{\alpha \beta} I_{\alpha}S_{\beta}.
\end{equation}Written explicitly, some of these matrices take the form
\begin{equation}
I_{z} = \sigma_z \otimes \Iden = \frac{1}{2} \left [ \begin{array}{cccc} 1 & 0 & 0 & 0 \\ 0 & 1 & 0 & 0 \\ 0 & 0 & -1 & 0 \\ 0 & 0 & 0 & -1 \end{array} \right ].
\end{equation}and
\begin{equation}
I_{z}S_z = \sigma_z \otimes \sigma_z = \frac{1}{4} \left [ \begin{array}{cccc} 1 & 0 & 0 & 0 \\ 0 & -1 & 0 & 0 \\ 0 & 0 & -1 & 0 \\ 0 & 0 & 0 & 1 \end{array} \right ].
\end{equation}The $15$ operators,
  $$ -i \{ I_\alpha, S_\beta, I_\alpha S_\beta \}, $$  for
$\alpha, \beta \in \{x, y, z \}$, form the basis for the Lie
algebra $\g = su(4)$, the $4 \times 4$, traceless skew Hermitian matrices.
For the coupled two spins, the generators $-iH_c, -iH_j \in su(4)$ and the evolution
operator $U(t)$ in Eq. (\ref{eq:sfcoupling}) is an element of $SU(4)$, the
$4 \times 4$, unitary matrices of determinant $1$.

The Lie algebra $\g = su(4)$ has a direct sum decomposition $ \g = \p \oplus \k $, where
\begin{equation}
\label{eq:su(4)}
\k = -i \{ I_\alpha , S_\beta \}, \ \ \p = -i \{ I_\alpha S_\beta \}.
\end{equation} Here $\k$ is a sub-algebra of $\g$ made from
local Hamiltonians and $\p$ nonlocal Hamiltonians. In Eq. \ref{eq:sfcoupling} , we have $-iH_j \in \k$ and $-iH_c \in \p$, It is easy to verify that
\begin{equation}
\label{eq:cartan}
[\k, \k] \subset \k , \ \ \ [\k, \p] \subset \p, \ \ [\p, \p] \subset \p.
\end{equation} This decomposition of a real semi-simple Lie algebra
$\g = \p \oplus \k$ satisfying (\ref{eq:cartan}) is called the
Cartan decomposition of the Lie algebra $\g$ \cite{Helg}.

This special structure of Cartan decomposition arising in dynamics of two coupled spins in Eq. \ref{eq:sfcoupling}, motivates study of a
general class of time optimal control problems.

Consider the following canonical problems. Given the evolution

\begin{equation}
\dot{U} = ( X_d + \sum_j u_j(t) X_j ) U, \ \ U(0) = \Iden,
\end{equation}
where $U \in SU(n)$, the special Unitary group (determinant $1$, $n \times n$ matrices $U$ such that $UU' = \Iden$, $'$ is conjugate transpose).
Where $$ X_d =  -i  \left[ \begin{array}{cccc} \lambda_1 & 0 & \hdots & 0 \\  0 & \lambda_2 & \hdots & 0 \\  \vdots & \vdots & \ddots & \vdots \\
0 & 0 & \hdots & \lambda_n \end{array} \right ], \ \ \sum \lambda_i = 0$$ and $\{X_j\}_{LA}$, the Lie algebra ($X_j$ and its matrix commutators) generated by generators $\{X_j \}$ is $\{X_j\}_{LA} = \k = so(n)$, skew symmetric matrices. We want to find the minimum time to steer this system between points of interest, assuming no bounds on our controls $u_j(t)$. Here again we have a Cartan decomposition on generators. Given $\g = su(n)$, traceless skew hermitian matrices , generators of $SU(n)$, we have $\g = \p \oplus \k$, where $\p = - i A$ where $A$ is traceless symmetric and $\k = so(n)$. As before, $X_d \in \p$ and $X_j \in \k$. We want to find time optimal ways to steer this system. We call this $\frac{SU(n)}{SO(n)}$ problem. 
We show for $n=4$, this system models the dynamics of two coupled nuclear spins in NMR spectrosocpy. 

Consider another problem evolving on $SU(2n)$.

\begin{equation}
\dot{U} = ( X_d + \sum_j u_j(t) X_j ) U,  \ \ U(0) = \Iden.
\end{equation}
Here $$ X_d =    \left[ \begin{array}{cccccc} 0 & \hdots & 0 & \lambda_1 & \hdots & 0  \\   \vdots  & \ddots & \vdots & \vdots & \ddots & \vdots \\  0 & \hdots & 0 & \hdots & 0 & \lambda_n \\  -\lambda_1 & \hdots & 0 & 0 & \hdots & 0 \\  \vdots  & \ddots & \vdots & \vdots & \ddots & \vdots \\ 0 & \hdots & -\lambda_n & 0 & \hdots & 0 \end{array} \right ]$$ and  $\{X_j\}_{LA} = \k = \left [ \begin{array}{cc} A  & 0 \\ 0 & B \end{array} \right ]$, space of block diagonal traceless skew Hermitian matrices. We want to find minimum time to steer this system between points of interest, assuming no bounds on our controls $u_j(t)$. Here again, we have a Cartan decomposition, of $\g = su(2n)$ as $\g = \p \oplus \k$ and $\p = \left [ \begin{array}{cc}  0 & Z \\ -Z' & 0 \end{array} \right ]$. $X_d \in \p$ and $X_j \in \k$, we want to find time optimal ways to steer this system. We call this $\frac{SU(2n)}{SU(n) \times SU(n) \times U(1)}$ problem.   
We show for $n=2$, this system models the dynamics of coupled electron-nuclear spin system in EPR \cite{Zeier}. 

In general, given $U$, in compact Lie group $G$ (such as $SU(n)$), with $X_d, X_j$ in its real semisimple (no abelian ideals) Lie algebra $\g$ and
\begin{equation}
\dot{U} = ( X_d + \sum_j u_j(t) X_j ) U,  \ \ U(0) = \Iden.
\end{equation} Given the cartan decomposition $\g = \p \oplus \k$, where $X_d \in \p$, $\{X_j\}_{LA} = \k$ and $K = \exp(\k)$ (product of exponentials of $\k$) a closed subgroup of G.  We want to find the minimum
time to steer this system between points of interest, assuming no
bounds on our controls $u_j(t)$. Since  $\{X_j\}_{LA} = \k$, any rotation
(evolution) in subgroup $K$ can be synthesized with evolution of $X_j$ \cite{Brockettc, Jurdjevic}. Since there are no bounds on $u_j(t)$, this can be done in arbitarily small time \cite{timeopt}. We call this $\frac{G}{K}$ problem.

The special structure of the problem aids in complete description of the reachable set. The elements of the reachable set at time $T$, takes the form
 $ U(T) \in$   
\begin{equation}
\label{eq:reachableintro}
S = K_1 \exp (T \sum_k \alpha_k \ \W_k X_d \W_k^{-1}) K_2, 
\end{equation} where $K_1, K_2, \W_k \in \exp(\k)$, and $\alpha_k > 0$, $\sum \alpha_k =1$ and $\W_k X_d \W_k^{-1}$ all commute and unbounded control suggests that $K_i, \W_k$ can be synthesized in negligible time.

This reachable set, which is formed from evolution of commuting Hamiltonians  $\W_k X_d \W_k^{-1}$, can be understood as follows. The Cartan decomposition of the Lie algebra $\g$, in Eq. (\ref{eq:cartan}) leads to a
decomposition of the Lie group $G$ \cite{Helg}. Let $\a$, denote the largest abelian sub-algebra contained
inside $\p$. Then any $X \in \p$ is $Ad_K$ conjugate to an element of
$\a$, i.e. $X = K a_1 K^{-1}$ for some $a_1 \in \a$.

Then, any arbitrary element of
the group $G$ can be written as
\begin{equation}
\label{eq:cartangroup}
G = K_0 \exp(X) = K_0 \exp(Ad_K(a_1)) = K_1 \exp(a_1) K_2,
\end{equation}for some $X \in \p$ where $K_i \in K$ and $a_1 \in
\a$. The first equation is a fact about geodesics in $G/K$ space \cite{Helg}, where $K = \exp(\k)$ is a closed subgroup of $G$.
Eq. (\ref{eq:cartangroup}) is called the KAK decomposition \cite{Helg}.

The results in this paper suggest that $K_1$ and $K_2$ can be
synthesized by unbounded controls $X_i$ in negligible time. The time
consuming part of the evolution $\exp(a_1)$ is synthesized by
evolution of Hamiltonian $X_d$. Time optimal strategy suggests
evolving $X_d$ and its conjugates $\W_k X_d
\W_k^{-1}$ where $\W_k X_d \W_k^{-1}$ all commute.

Written as evolution

$$ G = K_1 \prod_k \exp(t_k \W_k X_d \W_k^{-1}) \ K_2 $$
where $K_1, K_2, W_k$ take negligible time to synthesize using
unbounded controls $u_i$ and time-optimality is characterized by
synthesis of commuting Hamiltonians $\W_k X_d \W_k^{-1}$. This characterization of time
optimality, involving commuting Hamiltonians is derived using convexity ideas \cite{Kostant, timeopt}. The remaining
paper develops these notions.

The paper is orgaized as follows. In section 2, we study the $\frac{SU(n)}{SO(n)}$ problem.
In section 3, we study the  $\frac{SU(2n)}{SU(n) \times SU(n) \times U(1)}$ problem. 
In section 4, we study the general $\frac{G}{K}$ problem. We conclude in section 5 , with facts about 
roots and reflections, with application to dynamics of coupled spins.

Given Lie algebra $\g$, we use killing form $\langle x, y \rangle = tr(ad_xad_y)$ as a inner product on $\g$.
When $\g = su(n)$, we also use the inner product $\langle x, y \rangle = tr(x'y)$. We call this standard inner product.

\section{Time Optimal Control for $SU(n)/SO(n)$ problem}
\label{sec:second}

\begin{remark} {\rm Birkhoff convexity states, a real $n \times n$ matrix $A$ is doubly stochastic ($\sum_{i} A_{ij} = \sum_{j} A_{ij}= 1$, for $A_{ij} \geq 0$) iff it can be written as convex hull of permutation matrices $P_i$ (only one $1$ and everything else zero in every row and column). Given $\Theta \in SO(n)$ and $X =   \left[ \begin{array}{cccc} \lambda_1 & 0 & \hdots & 0 \\  0 & \lambda_2 & \hdots & 0 \\  \vdots & \vdots & \ddots & \vdots \\
0 & 0 & \hdots & \lambda_n \end{array} \right ]$, we have $diag(\Theta X \Theta^T) = B \ diag(X)$ where $diag(X)$ is a column vector containing diagonal entries of $X$ and $B_{ij} = (\Theta^{ij})^2$ and hence $B$ is a doubly stochastic matrix which can be written as convex sum of permutations. Therefore $B \ diag(X) = \sum_i \alpha_i P_i\ diag(X)$, i.e. diagonal of a symmetric matrix $\Theta X \Theta^T$, lies in convex hull of its eigenvalues and its permutations. This is called Schur convexity. }\end{remark}

We now give an elementary proof of special case of the KAK decomposition in Eq. (\ref{eq:cartangroup}), where $G = SU(n)$ has a closed subgroup $K = SO(n)$ and a Cartan decompostion of its Lie algerbra $\g = su(n)$ as $\g = \p  \oplus \k$, for $\k = so(n)$ and $p = -i A$ where $A$ is traceless symmetric and $\a$ 
is maximal abelian subalgebra of $\p$ , such that $\a = -i \left[ \begin{array}{ccc} \lambda_1 & \dots & 0  \\  0 & \ddots & 0 \\  0 & 0 & \lambda_n  \end{array} \right ],$ where $\sum_i \lambda_i = 0$.

\begin{theorem}\label{th:SUdecompose}{\rm Let $U \in SU(n)$, then $U = \Theta_1 \exp(\Omega)\Theta_2$ where $\Theta_1, \Theta_2 \in SO(n)$ and
$$\Omega = -i \left[ \begin{array}{ccc} \lambda_1 & \dots & 0  \\  0 & \ddots & 0 \\  0 & 0 & \lambda_n  \end{array} \right ], $$ where $\sum_i \lambda_i = 0$.}
\end{theorem}

Observe $UU^T$ is in $SU(n)$. The eigenvalues of $UU^T$ are of the form $\exp(j \theta)$.

$$ UU^T z = \exp(j \theta) z. $$

$$ \exp(-j \frac{\theta}{2}) U^T z = \exp(j \frac{\theta}{2}) (U^T)^{\ast} z.  $$

$$ (C + iD) z = (C - iD)z. $$

$$ D (x + iy) = 0. $$

This implies $UU^T x = \exp(j \theta) x $ and $UU^T y = \exp(j \theta) y $. This implies $ UU^T = \Theta \Sigma \Theta'$, where columns of $\Theta$ are real, perpendicular, and

$$ \Sigma = \left[ \begin{array}{ccc} \exp(-i \lambda_1) & \dots & 0  \\  0 & \ddots & 0 \\  0 & 0 & \exp(-i \lambda_n)  \end{array} \right ] $$

where $\Sigma \in SU(n)$. Let $ U = \Theta \Sigma^{\frac{1}{2}} V$. $ UU^T = \Theta \Sigma \Theta' = \Theta \Sigma^{\frac{1}{2}} VV^T \Sigma^{\frac{1}{2}} \Theta' $.

Implying $VV^T = \Iden$. Then $U = \Theta \Sigma^{\frac{1}{2}} V$, where $\Theta, V$ can be chosen in $SO(n)$ and 

$$ \Sigma^{\frac{1}{2}} = \left[ \begin{array}{ccc} \exp(-i \mu_1) & \dots & 0  \\  0 & \ddots & 0 \\  0 & 0 & \exp(-i \mu_n)  \end{array} \right ], $$

where  $ \sum \mu_i = 2 m \pi $. Choose $\mu_n \rightarrow \mu_n - 2 m \pi$ so that  $ \sum \lambda_i = 0$ and result follows.{\bf q.e.d}

\n We now give a proof of the reachable set in Eq. (\ref{eq:reachableintro}), for the $\frac{SU(n)}{SO(n)}$ problem.

\begin{theorem}\label{th:model1}{ \rm Let $P(t) \in SU(n)$ be a solution to the differential equation $$ \dot{P} = Ad_{K(t)}(X_d) P, $$ where $Ad_K(X_d) = K X K^{-1}$ and
$ X_d = -i  \left[ \begin{array}{cccc} \lambda_1 & 0 & \hdots & 0 \\  0 & \lambda_2 & \hdots & 0 \\  \vdots & \vdots & \ddots & \vdots \\
0 & 0 & \hdots & \lambda_n \end{array} \right ]$. The elements of the reachable set at time $T$, take the form $ K_1 \exp(-i \mu T) K_2 $, where $K_1, K_2 \in SO(n)$ and $\mu \prec \lambda$ ($\mu$ lies in convex hull of $\lambda$ and its permutations), where $\lambda = (\lambda_1, \dots, \lambda_n)'$.}\end{theorem}

{\bf Proof:} As a first step, discretize the evolution of $P(t)$, as piecewise constant evolution,

\begin{equation}
\label{eq:discretize}
P_n = \prod_i \exp(Ad_{k_i} (X_d) \tau), 
\end{equation}
of steps of size $\tau$. For arbitrary $t \in [0, T]$ we look at the evolution of $P(t)$. Let 
$t \in [(n-1) \tau, n \tau]$. Choose small step $\Delta$, such that $n \tau - t < \Delta$, then
$P(t + \Delta) = \exp(Ad_K(X_d)\Delta) P(t)$.

\n From theorem \ref{th:SUdecompose}, $P(t) = K_1 \left [ \begin{array}{cccc} \exp(i \phi_1) & 0 & 0 & 0 \\ 0 & \exp(i \phi_2) & 0 & 0 \\ 0 & 0 & \ddots & 0  \\ 0 & 0 & 0 & \exp(i \phi_n) \end{array} \right ] K_2 $, where $K_1, K_2 \in SO(n)$,where to begin with, assume eigenvalues $\phi_j - \phi_k \neq n \pi$, where $n$ is an integer.
Let $ K_1(t + \Delta) = \exp(\Omega_1 \Delta) K_1(t)$, $K_2(t + \Delta) = \exp(\Omega_2 \Delta) K_2 $, and $A(t + \Delta) = \exp(a \Delta) A(t)$, where, $\Delta$, $\Omega_1$, $\Omega_2$ and $a$ are detailed below. Let $Q(t + \Delta) = K_1(t + \Delta) A(t + \Delta) K_2(t + \Delta)$.
\begin{equation}
\label{eq:1}
Q(t + \Delta) = \exp(\Omega_1 \Delta)\exp(K_1 a K_1' \Delta)\exp(K_1 A \Omega_2 A' K_1' \Delta) P(t).
\end{equation}

Let

\begin{equation}
\label{eq:2}
P(t + \Delta) =  \exp(Ad_{K}(H_d) \Delta  ) P(t). 
\end{equation} 

We equate $P(t + \Delta)$ and $Q(t+ \Delta)$ to first order in $\Delta$. This gives,

\begin{equation}
\label{eq:solve1}
Ad_{K}(X_d) = \Omega_1 + K_1 a K_1' + K_1 A \Omega_2 A' K_1'.
\end{equation}

Multiplying both sides with $K_1'(\cdot)K_1$ gives

\begin{equation}
\label{eq:solve2}
Ad_{\bar K}(X_d) = \Omega_1' + a  + A \Omega_2 A'.
\end{equation}
where,  $\bar K = K_1' K$ and $\Omega_1' = K' \Omega K $.

We evaluate $ A \Omega_2 A^{\dagger} $, for $\Omega_2 \in so(n)$.

$$ D = \{ A  \Omega_2 A^{\dagger} \}_{kl} = \exp \{i(\phi_k - \phi_l)\} (\Omega_2)_{kl} = \cos (\phi_k - \phi_l) (\Omega_2)_{kl}  + i \underbrace{\sin(\phi_k - \phi_l) (\Omega_2)_{kl}}_{R_{kl}} $$
such that $R$ is traceless symmetric matrix with $P_1 = i R \in \p$ and onto
$\a^{\perp}$, by appropriate choice of $\Omega_2$.

Given $Ad_{\bar K}(X_d) \in \p$, we decompose it as

$$ P ( Ad_{\bar K}(X_d)) +  Ad_{\bar K}(X_d)^{\perp}, $$ with $P$ denoting the projection onto $\a$ ( $\a = -i \left[ \begin{array}{ccc} \lambda_1 & \dots & 0  \\  0 & \ddots & 0 \\  0 & 0 & \lambda_n  \end{array} \right ],$ where $\sum_i \lambda_i = 0$.) w.r.t to standard inner product and $Ad_{\bar K}(X_d)^{\perp}$ to the orthogonal component. If $\phi_i - \phi_j \neq 0, \pi$, we can solve for $(\Omega_2)_{ij}$ such that
$P_1 = Ad_{\bar K}(X_d)^{\perp}$. This gives $\Omega_2$. Let $a = P(Ad_{\bar K}(X_d))$.

As described above in Eq. (\ref{eq:solve2}), we choose $\Omega_1' = Ad_{\bar{K}}(X_d)^{\perp} - A \Omega A^{\dagger} \in \k $.

Then

$$P(t + \Delta) - Q(t + \Delta) = o(\Delta^2). $$

Consider the case, when $A$ is degenerate. Let,
\begin{equation}\label{eq:blocks}
A = \left[ \begin{array}{cccc} A_1 & 0 & \hdots & 0 \\  0 & A_2 & \hdots & 0 \\  \vdots & \vdots & \ddots & \vdots \\
0 & 0 & \hdots & A_n \end{array} \right ],\end{equation}
where $A_k$ is $n_k$ fold degenerate ( modulo sign) described by $n_k \times n_k$ block. WLOG, we arrange

$$ A_k = \left[ \begin{array}{cccccc} \exp(i \phi_k) & 0 & \hdots & \hdots & \hdots & 0\\  \vdots & \ddots & \vdots & \vdots & \vdots & 0 \\ 0 & \hdots & \exp(i \phi_k) & \hdots & \dots & 0 \\  0 & \vdots & \hdots & -\exp(i \phi_k) & \hdots & \vdots  \\ \vdots & \vdots & \hdots & \vdots & \ddots & \vdots \\  0 & \hdots & \hdots & \hdots & 0 & -\exp(i \phi_k) \end{array} \right ].  $$

Consider the decomposition

$$  Ad_{\bar K}(X_d) = P(Ad_{\bar K}(X_d)) +  Ad_{\bar K}(X_d)^{\perp}, $$ where $P$ denotes projection onto $n_k \times n_k$ blocks in equation \ref{eq:blocks} and  $Ad_K(X_d)^{\perp}$, the orthogonal complement.

\begin{equation} P( \left[ \begin{array}{cccc} X_{11} & X_{12} & \hdots & X_{1n} \\  X_{21} & X_{22} & \hdots & X_{2n} \\  \vdots & \vdots & \ddots & \vdots \\
X_{n1} & X_{n2} & \hdots & X_{nn} \end{array} \right ]) =  \left[ \begin{array}{cccc} X_{11} & 0 & \hdots & 0 \\  0  & X_{22}  & \hdots & 0  \\  \vdots & \vdots & \ddots & \vdots \\
0 & 0 & \hdots & X_{nn} \end{array} \right ] ,\end{equation} where $X_{ij}$ are blocks.

We can solve for $(\Omega_2)_{ij}$ such that $P_1 = Ad_{\bar K}(X_d)^{\perp}$. This gives $\Omega_2$ in Eq. (\ref{eq:solve2}).

Choose, $ Ad_{\bar{K}}(X_d)^{\perp} - A \Omega A^{\dagger} = \Omega_1' \in \k $, and  Let $H_1 = \exp(h_1)$ be a rotation formed from block diagonal matrix

\begin{equation}\label{eq:rotblocks}
H_1 = \left[ \begin{array}{cccc} \Theta_1 & 0 & \hdots & 0 \\  0 & \Theta_2 & \hdots & 0 \\  \vdots & \vdots & \ddots & \vdots \\
0 & 0 & \hdots & \Theta_n \end{array} \right ],\end{equation}

where $\Theta_k$ is $n_k \times n_k$ sub-block in $SO(n_k)$.
$H_1 = exp(h_1)$ is chosen such that
$$ H_1' P ( Ad_{\bar K}(X_d)) H_1 = a$$ is a diagonal matrix.
$H_2 = \exp(\underbrace{A^{-1}h_1A}_{h_2})$,
where $h_2$ is skew symmetric, such that

\begin{equation}\label{eq:rotgenblocks}
h_1 = \left[ \begin{array}{cccc} \theta_1 & 0 & \hdots & 0 \\  0 & \theta_2 & \hdots & 0 \\  \vdots & \vdots & \ddots & \vdots \\
0 & 0 & \hdots & \theta_n \end{array} \right ], h_2 = \left[ \begin{array}{cccc} \hat \theta_1 & 0 & \hdots & 0 \\  0 & \hat \theta_2 & \hdots & 0 \\  \vdots & \vdots & \ddots & \vdots \\
0 & 0 & \hdots & \hat \theta_n \end{array} \right ], \end{equation} where

$\theta_k, \hat \theta_k$ is $n_k \times n_k$ sub-block in $so(n_k)$, related by

\begin{equation}\label{eq:rotgenblocks1}
\hat \theta_k = A_k ' \theta_k A_k ,\  \theta_k = \left[ \begin{array}{cc} \theta_{11} & \theta_{12} \\  - \theta_{12}^{\dagger}  & \theta_{22} \end{array} \right ],  \hat \theta_k = \left[ \begin{array}{cc} \theta_{11} & -\theta_{12} \\  \theta_{12}^{\dagger}  & \theta_{22} \end{array} \right ] \end{equation}

Note $H_1' P ( Ad_k(X_d)) H_1 = a$ lies in convex hull of eigenvalues of $X_d$. This is true if we look at the diagonal of
$H_1' Ad_K(X_d) H_1$, it follows from Schur Convexity. The diagonal of $H_1' Ad_k(X_d)^\perp H_1 $ is zero as its inner product

$$ tr (a_1 H_1' Ad_k(X_d)^\perp H_1) = tr (H_1 a_1 H_1' Ad_k(X_d)^\perp) = 0. $$ as $H_1 a_1 H_1'$ has block diagonal form which is perpendicular to $Ad_k(X_d)^\perp$. Therefore diagonal of  $H_1' P ( Ad_k(X_d)) H_1$ is same as diagonal of $H_1' Ad_K(X_d) H_1$.

Let

$$ Q(t + \Delta) = \exp(\Omega_1 \Delta) K_1 \exp(P(Ad_{\bar K}(X_d) \Delta)) H_1 A H_2^{\dagger} \exp(\Omega_2 \Delta) K_2. $$

\begin{equation}
\label{eq:defa}
Q(t + \Delta) = \exp(\Omega_1 \Delta) K_1 H_1 \exp(a \Delta) A H_2^{\dagger} \exp(\Omega_2 \Delta) K_2. 
\end{equation}

where the above expression can be written as

$$ Q(t + \Delta) = \exp(\Omega_1 \Delta) \exp(K_1 H_1 a H_1'K_1' \Delta) \exp(K_1 A \Omega_2 A' K_1' \Delta) P(t). $$

Where $\Omega_1$, $H_1, a, \Omega_2$, are chosen such that

$$ (\Omega_1 + K_1 H_1 a H_1' K_1' + K_1 A \Omega_2 A' K_1') = Ad_K(X_d). $$

$$ (\Omega_1' + H_1 a H_1' + A \Omega_2 A') = Ad_{\bar K}(X_d). $$

$$ Q(t + \Delta)- P(t + \Delta) = o(\Delta^2) P(t). $$

$$ Q(t + \Delta) = (I + o(\Delta^2))P(t + \Delta). $$

$$ Q(t + \Delta)Q(t + \Delta)^{T} = (I + o(\Delta^2))P(t + \Delta)P^{T}(t + \Delta) (I + o(\Delta^2)) = P(t+ \Delta)P^{T}(t + \Delta) [ I + o(\Delta^2)]. $$

$$ P(t+\Delta)P^{T}(t+ \Delta) = K_1 \left[ \begin{array}{cccc} \exp(i 2 \phi_1) & 0 & \hdots & 0 \\  0 & \exp(i 2 \phi_2) & \hdots & 0 \\  \vdots & \vdots
 & \ddots & \vdots \\
 0 & 0 & \hdots & \exp(i 2 \phi_n) \end{array} \right ] K_1^{T}. $$

Let $F = P(t + \Delta)P^T(t + \Delta)$ and $G = Q(t + \Delta)Q^T(T + \Delta)$ we relate the eigenvalues, of $F$ and $G$. Given $F, G$, as above, with $| F - G | \leq \epsilon$, and a ordered set of eigenvalues of F, denote $\lambda(F) = \left[ \begin{array}{c} \exp(i 2 \phi_1) \\  \exp(i 2 \phi_2) \\  \vdots \\ \exp(i 2 \phi_n) \end{array} \right ]$, there exists an ordering (correspondence) of eigenvalues of $G$, such that $|\lambda(F) - \lambda(G)| < \epsilon$.

Choose an ordering of $\lambda(G)$ call $\mu$ that minimizes $|\lambda(F) - \lambda(G)|$.

$F = U_1 D(\lambda)U_1'$ and $G = U_2 D(\mu) U_2'$ , where $D(\lambda)$ is diagonal with diagonal as $\lambda$, let $U = U_1'U_2$,

$$ | F- G|^2 = |D(\lambda) - U D(\mu) U' |^2 = |\lambda|^2  + |\mu|^2  -
tr(D(\lambda)' U D(\mu) U' + (U D(\mu) U)' D(\lambda)), $$

By Schur convexity,

$$ tr(D(\lambda)' U D(\mu) U' + (U D(\mu) U')' D(\lambda)) = \sum_i \alpha_i (\lambda' P_i(\mu) + P_i(\mu)' \lambda), $$
where $P_i$ are permutations. Therefore $|F-G|^2 > |\lambda - \mu|^2$.

Therefore,

$$ \lambda(QQ^T(t+\Delta)) = \lambda(PP^{T}(t+\Delta)) + o(\Delta^2). $$

The difference $$ o(\Delta^2)  = \underbrace{\exp ((\Omega_1 + K_1 H_1 a H_1' K_1' + K_1 A \Omega_2 A' K_1')\Delta)}_{\exp(Ad_K(X_d)\Delta)} - \exp (\Omega_1 \Delta) \exp(K_1 H_1 a H_1' K_1' \Delta) \exp (K_1 A \Omega_2 A' K_1' \Delta), $$

is regulated by size of $\Omega_2$, which is bounded by $|\Omega_2| \leq \frac{\|X_d \|}{\sin(\phi_i -\phi_j)}$, where $\sin(\phi_i-\phi_j)$ is smallest non-zero difference. $\Delta$ is chosen small enough such that $|o(\Delta^2)| < \epsilon \Delta$.

For each point $t \in [ 0, T]$, we choose a open nghd $N(t) = (t - N_t, t + N_t )$ ( $[0, N_0)$ and $(T-N_T, T]$ ), such that

$o_t(\Delta^2) < \epsilon \Delta$ for $\Delta \in N(t)$. $N(t)$ forms a cover of
$[0, T]$. We can choose a finite sub-cover. Consider trajectory at points $(P(t_1), P(t_2), \dots P(t_n))$.
Let $t_{i, i+1}$ be the point in intersection of $N(t_i)$ and $N(t_{i+1})$. Let $\Delta_i^{+} = t_{i, i+1}-t_i$ and $\Delta_{i+1}^{-} = t_{i+1} - t_{i, i+1}$. We consider points
$P(t_i), P(t_{i+1}), P(t_{i, i+1}), \underbrace{Q(t_{i} + \Delta_i^{+})}_{Q_{i+}}, \underbrace{Q(t_{i+1}-\Delta_{i+1}^{-})}_{Q_{(i+1)-}}$

\begin{figure}[htb]
\begin{center}
\includegraphics[scale=.4]{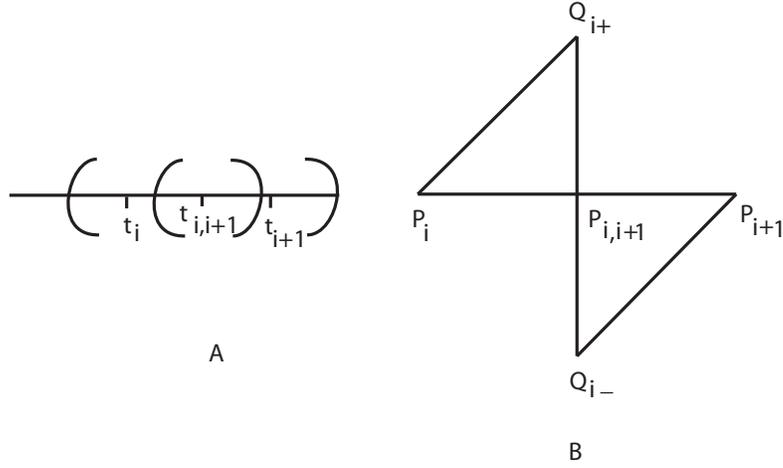}
\caption{Figure A shows collection of overlappings neighbourhoods forming the finite subcover. Figure B depicts 
$P_i$, $P_{i+1}$, $Q_{i+}$, $Q_{i-}$, $P_{i, i+1}$ as in proof of theorem \ref{th:model1}.}
\end{center}
\end{figure}

The recursive relation gives,

\begin{eqnarray}
\lambda(Q_{i+}Q_{i+}^{T}) &=& \exp(2 a_i^{+} \Delta_i^+)\ \lambda(P_{i}P_{i}^{T})  \\
\lambda(P_{i, i+1}P_{i, i+1}^{T}) &=& \lambda(Q_{i+}Q^{T}_{i+}) + o((\Delta_i^+)^2) \\
\lambda(Q_{(i+1)-}Q_{(i+1)-}^{T}) &=& \lambda(P_{i, i+1}P_{i, i+1}^{T}) + o((\Delta^{-}_{i+1})^2) \\
\exp(-2 a_{i+1}^{-} \Delta_{i+1}^{-})\ \lambda(P_{i+1}P_{i+1}^T) &=& \lambda(Q_{(i+1)-}Q^{T}_{(i+1)-})
\end{eqnarray}where $a_i^+$ and $a_{i+1}^-$ correspond to $a$ in Eq. (\ref{eq:defa}) and lie in the convex hull of the eigenvalues $X_d$.

Adding the above equations,

\begin{equation}
\lambda(P_{i+1}P_{i+1}^{T}) = \exp (o(\Delta^2))\ \exp(2 ( a_i^+ \Delta_i^+ + a_{i+1}^{-}\Delta_{i+1}^{-})\ \lambda(P_i P_i^{\dagger}).
\end{equation}where $o(\Delta^2)$ is diagonal.

\begin{equation}
\lambda(P_nP_n^{T}) =  \exp(\underbrace{\sum o(\Delta^2)}_{\leq \epsilon T})\ \exp(2 \sum_i a_i^{+} \Delta_i^{+} + a_{i+1}^{-}\Delta_{i+1}^{-})\ \lambda(P_1P_1^{T}).
\end{equation}

where $|\exp(i \alpha) -1| = 2 \sin\frac{|\alpha_1|}{2} > \frac{|\alpha_1|}{2}$. Therefore, $|\exp(i \alpha) - 1| \leq \frac{\theta}{2}$ implies $|\alpha| \leq \theta$.

\begin{equation}
\lambda(P_nP_n^{T}) = \exp(\underbrace{\sum o(\Delta^2)}_{\leq \epsilon T})\ \exp(2 T \sum_k \alpha_k P_k(\lambda))\ \lambda(P_1P_1^{T}) = \exp(\underbrace{\sum o(\Delta^2)}_{\leq \epsilon T})\ \exp(2 \mu T)\ \lambda(P_1P_1^{T}),
\end{equation}where $\mu \prec \lambda$ and $P_1 = I$.

\begin{equation}
P_n = K_1 \exp(\mu T) \exp(\frac{1}{2}\underbrace{\sum o(\Delta^2)}_{\leq \epsilon T}) K_2.
\end{equation} Note, $|P_n - K_1 \exp(\mu T) K_2| = o(\epsilon)$. This implies that $P_n$ belongs to the compact set
$K_1 \exp(\mu T) K_2$, else it has minimum distance from this compact set and by making $\Delta \rightarrow 0$ and hence $\epsilon \rightarrow 0$, we can make this arbitrarily small.
In Eq. \ref{eq:discretize}, $P_n \rightarrow P(T)$ as $\tau \rightarrow 0$. Hence $P(T)$ belongs to compact set $K_1 \exp(\mu T) K_2$. {\bf q.e.d}

\begin{corollary}
\label{cor:model1}{ \rm Let $U(t) \in SU(n)$ be a solution to the differential equation $$ \dot{U} = (X_d + \sum_i u_i X_i )U, $$ where $\{X_i\}_{LA}$, the Lie algebra generated by $X_i$, is $so(n)$ and
$ X_d = -i  \left[ \begin{array}{cccc} \lambda_1 & 0 & \hdots & 0 \\  0 & \lambda_2 & \hdots & 0 \\  \vdots & \vdots & \ddots & \vdots \\
0 & 0 & \hdots & \lambda_n \end{array} \right ]$. The elements of reachable set at time $T$, takes the form $ U(T) \in K_1 \exp(-i \mu T) K_2 $, where $K_1, K_2 \in SO(n)$ and $\mu \prec \lambda$, where $\lambda = (\lambda_1, \dots, \lambda_n)'$ and the set  $S = K_1 \exp(-i \mu T) K_2$ belongs to the closure of reachable set. }\end{corollary}

\no {\bf Proof:} Let $V(t) = K'(t) U(t) $, where, $\dot{K} = (\sum_i u_i X_i) K$. Then
$$\dot{V}(t) = Ad_{K'(t)}(X_d) V(t). $$ From theorem \ref{th:model1}, we have $V(T) \in K_1 \exp(-i \mu T) K_2$. Therefore $U(T) \in  K_1 \exp(-i \mu T) K_2$. Given
$$ U = K_1 \exp(-i \mu T) K_2 =  K_1 \exp(-i \sum_j \alpha_j P_j(\lambda) T) K_2 = K_1 \prod_j \exp(-i t_j X_d ) K_j,\ \ \sum t_j = T  .$$ We can synthesize $K_j$ in negligible time, therefore $| U(T) - U | < \epsilon$, for any desired $\epsilon$. Hence $U$ is in closure of reachable set. {\bf q.e.d}

\begin{remark}{\rm We now show how theorem \ref{th:SUdecompose} and \ref{th:model1} can be mapped to results on decomposition and reachable set for coupled spins/qubits. 
Consider the transformation

$$ W = \exp(-i \pi I_yS_y) \exp(-i \frac{\pi}{2} I_z )$$

The transformation maps the algebra $\k = su(2) \times su(2) = \{ I_\alpha, S_\alpha \}$ to $\k_1 = so(4)$, four dimensional skew symmetric matrices, i.e., $Ad_W(\k) = \k_1$.
The transformation maps $\p = \{ I_\alpha S_\beta \}$ to $\p_1 = -i A $, where $A$ is traceless symmetric and maps $\a = -i \{ I_xS_x, I_yS_y, I_zS_z  \}$ to $\a_1 = -i\{-\frac{S_z}{2}, \frac{I_z}{2}, I_zS_z \}$, space of diagonal matrices in $\p_1$, such that the triplet $(a_x, a_y, a_z)$ gets mapped to the four vector (the diagonal) $(\lambda_1, \lambda_2, \lambda_3, \lambda_4) = (a_y + a_z - a_x, a_x + a_y - a_z, -(a_x + a_y + a_z), a_x + a_z - a_y)$.}
\end{remark}

\begin{corollary}{\rm {\bf Canonical Decomposition:} Given the decomposition of SU(4) from theorem \ref{th:SUdecompose}, we can write 
$$ U = \exp(\Omega_1) \exp( -i \left[ \begin{array}{ccc} \lambda_1 & \dots & 0  \\  0 & \ddots & 0 \\  0 & 0 & \lambda_4  \end{array} \right ]) \exp(\Omega_2),$$
where $\Omega_1, \Omega_2 \in so(4)$. We write above as 
$$ U = \exp(\Omega_1) \exp( -i (-\frac{a_x}{2} S_z  + \frac{a_y}{2} I_z  + a_z I_zS_z)) \exp(\Omega_2),$$ Multiplying both sides with $W'(.)W$ gives
$$ W' U W = K_1 \exp(-i a_x I_xS_x + a_y I_yS_y + a_z I_zS_z) K_2, $$ where $K_1, K_2 \in SU(2) \times SU(2)$ local unitaries and we can rotate to $a_x \geq a_y \geq |a_z|$.}
\end{corollary}

\begin{corollary}{\rm {\bf Digonalization} Given $-iH_c = -i \sum_{\alpha \beta} J_{\alpha \beta}I_\alpha S_\beta$ , there exists a local unitary $K$ such that 
$$ K (-iH_c) K' =  -i(a_x I_xS_x + a_y I_yS_y + a_z I_zS_z), a_x \geq a_y \geq |a_z|. $$ Note $W (-iH_c)W' \in \p_1$. Then choose $\Theta \in SO(n)$ such that
$\Theta W (-iH_c)W' \Theta' =  -i (-\frac{a_x}{2} S_z  + \frac{a_y}{2} I_z  + a_z I_zS_z)$ and hence 
$$ (W' \exp(\Omega) W) (-i H_c ) (W \exp(\Omega) W')' =   -i(a_x I_xS_x + a_y I_yS_y + a_z I_zS_z). $$ Where $ K = W' \exp(\Omega) W$ is a local unitary. 
We can rotate to ensure $a_x \geq a_y \geq |a_z|$.}\end{corollary}

\begin{corollary}{\rm Given the evolution of coupled qubits $\dot U = -i(H_c + \sum_j u_j H_j) U$, we can diagonalize 
$H_c = \sum_{\alpha \beta} J_{\alpha \beta} I_{\alpha}S_{\beta}$  by local unitary $X_d = K' H_c K = a_x I_xS_x + a_y I_yS_y + a_z I_zS_z$, $a_x \geq a_y \geq |a_z|$, which we write as 
triple $(a_x, a_y, a_z)$ . From this, there are 24 triples obtained by permuting and changing sign of any two by local unitary. Then $U(T) \in S$ where 
$$ S = K_1 \exp(T \sum_i \alpha_i (a_i, b_i, c_i)) K_2, \ \alpha_i > 0 \ \sum_{i}\alpha_i = 1. $$ Furthermore $S$ belongs to the closure of the reachable set.
Alternate description of $S$ is 
$$U = K_1 \exp( -i ( \alpha I_xS_x + \beta I_yS_y  + \gamma I_zS_z))K_2,  \ \ \alpha \geq \beta \geq |\gamma|, $$
$\alpha \leq a_x T$ and $\alpha + \beta \pm \gamma \leq (a_x + a_y \pm a_z) T$.} 
\end{corollary}

\no {\bf Proof:} Let $V(t) = K'(t) U(t) $, where, $\dot{K} = (-i \sum_j u_j X_j) K$. Then
$$\dot{V}(t) = Ad_{K'(t)}(-iX_d) V(t). $$ 
Consider the product $$ V = \prod_i \exp(Ad_{K_i}(-i X_d) \Delta t) $$ where $K_i \in SU(2)\otimes SU(2)$ and $X_d = a_x I_xS_x + a_y I_yS_y  + a_z I_zS_z$, where $a_x \geq a_y \geq |a_z|$. Then,

$$ W V W' =  \prod_i \exp(Ad_{WK_iW'}(-i WX_dW') \Delta t) $$
Observe $WK_iW' \in SO(4)$ and  $WX_dW'= diag(\lambda_1, \lambda_2, \dots , \lambda_4)$. Then using results from theorem \ref{th:model1}, we have

$$ WVW' = J_1 \exp(-i \mu)J_2  = J_1 \exp(-i \sum_j \alpha_j P_j(\lambda)) J_2 , \ \ J_1, J_2 \in SO(4), \ \ \mu \prec \lambda T $$

Multiplying both sides with $W'(\cdot)W$ , we get

$$ V = K_1 \exp(T \sum_i \alpha_i (a_i, b_i, c_i)) K_2, \ \alpha_i > 0 \ \sum_{i}\alpha_i = 1. $$

which we can write as

$$ V = K_1 \exp ( -i (\alpha I_xS_x + \beta I_yS_y + \gamma I_zS_z)) K_2, \ \ \alpha \geq \beta \geq |\gamma|, $$ where
using $\mu \prec \lambda T $, we get,

\begin{eqnarray}
\alpha + \beta - \gamma &\leq& (a_x + a_y - a_z) T \\
\alpha &\leq& a_x T  \\
\alpha + \beta + \gamma &\leq& (a_x + a_y + a_z) T.
\end{eqnarray} Furthermore $U = KV$. Hence the proof. {\bf q.e.d}

\section{Time Optimal Control for $\frac{SU(2n)}{SU(n) \times SU(n) \times U(1)}$ problem}
\label{sec:third}

\begin{remark}\label{rem:stabilizer}{\rm {\bf Stabilizer:} Let $\g = \p \oplus \k$ be cartan decomposition of real semisimple Lie algebra $\g$ and $\a \in \p $ be its Cartan subalgebra. Let $a \in \a$. $ad_a^2 : \p \rightarrow \p$ is symmetric in basis orthonormal wrt to the killing form. We can diagonalize $ad_a^2$. Let $Y_i$ be eigenvectors with nonzero (negative) eigenvalues  $-\lambda_i^2$. Let $X_i = \frac{[a, Y_i]}{\lambda_i}$, $\lambda_i > 0$.
$$ ad_a(Y_i) = \lambda_i X_i, \ \ ad_a(X_i) = -\lambda_i Y_i. $$

$X_i$ are independent, as $\sum \alpha_i X_i = 0$ implies $- \sum \alpha_i \lambda_i Y_i=0$. Since $Y_i$ are independent, $X_i$ are independent. Given $X \perp X_i$ ,
then $[a, X ] =0$, otherwise we can decompose it in eigenvectors of $ad_a^2$, i.e.,
$[a, X] = \sum_i \alpha_i a_i + \sum_j \beta_j Y_j $, where $a_i$ are zero eigenvectors of $ad_a^2$. Since $0 = \langle X [a[a, X] \rangle = - \| [a, X] \|^2$, which means $[a, X] =0$. This is a contradiction. $Y_i$ are orthogonal, implies $X_i$ are orthogonal, $\langle [a, Y_i] [a, Y_j] \rangle =  \langle [a, [a, Y_i] Y_j \rangle = \lambda_i^2 \langle Y_i Y_j \rangle = 0$.
Let $\k_0 \in \k$ satisfy $[a, \k_0] = 0$. Then $\k_0 = \{ X_i \}^{\perp}$.

$\tilde Y_i$ denote eigenvectors that have $\lambda_i$ as non-zero integral multiples of $\pi$. $\tilde{X}_i$ are $ad_a$ related to $\tilde{Y}_i$. We now reserve $Y_i$ for non zero eigenvectors that are not integral multiples of $\pi$.

Let
$$\f = \{a_i\} \oplus \tilde{Y}_i, \ \ \ \h = \k_0 \oplus \tilde{X}_i, $$

$\tilde{X_i}, X_l, k_j$ where $k_j$ forms a basis of $\k_0$, forms a basis of $\k$. Let $A = \exp(a)$.

$A k A^{-} = A (\sum_i \alpha_i X_i + \sum_l \alpha_l \tilde{X}_l + \sum_j \alpha_j k_j ) A^{-} $, where $k \in \k$

$A k A^{-} = \sum_i \alpha_i [ \cos(\lambda_i) X_i - \sin(\lambda_i) Y_i ] + \sum_l \pm \alpha_l \tilde{X}_l + \sum_j \alpha_j k_j  $

The range of $A (\cdot) A^{-}$ in $\p$, is perpendicular to $\f$. Given $Y \in \p$ such that $Y \in \f^{\perp}$. The norm $\|X\|$ of $X \in \k$,
such that $\p$ part of $A X A^{-1}|_{\p} = Y$ satisfies
\begin{equation}
\label{eq:nghdbnd}
\| X \| \leq \frac{\|Y \|}{\sin \lambda_s}.
\end{equation} 
where $\lambda_{s}^2$ is the smallest nonzero eigenvalue of $-ad_a^2$ such that $\lambda_s$ is not an integral multiple of $\pi$.

$A^2 k A^{-2}$ stabilizes $\h \in \k$ and $\f \in \p$. If $k \in \k$, is stabilized by $A^2 (\cdot) A^{-2}$,
$\lambda_i = n \pi$, i.e., $k \in \h$. This means $\h$ is an sub-algebra, as the Lie
bracket of $[y, z] \in \k$ for $y, z \in \h$ is stabilized by $A^2 (\cdot) A^{-2}$.

Let $H = \exp(\h)$, be an integral manifold of $\h$. Let $\tilde{H} \in K$ be the solution
to $A^2 \tilde H A^{-2} = \tilde H$ or $A^2 \tilde H - \tilde H A^{-2} = 0$. $\tilde{H}$ is closed, $H \in \tilde H$. We show that
$\tilde H$ is a manifold. Given $H_0 \in \tilde{H} \in K$, where $K$ is closed,
we have a $\exp(B_{\delta}^{\k})$ nghd of $H_0$, in $\exp(B_{\delta})$ ball nghd of $H_0$, which is one to one. For $x \in B_{\delta}^{\k}$,
$ A^2 \exp(x) A^{-2} = \exp(x)$, implies,

\begin{equation}
A^2 \exp(\sum_i \alpha_i X_i + \sum_l \beta_l \tilde{X}_l + \sum_j \gamma_j k_j )H_0 A^{-2} = 
\exp (\sum_i \alpha_i \cos(2 \lambda_i) X_i - \sin(2 \lambda_i) Y_i + \sum_l \beta_l \tilde{X}_l + \sum_j \gamma_j k_j )H_0,
\end{equation} then by one to one, $\exp(B_{\delta})$, we get $\alpha_i = 0$ and $x \in \h$. Therefore $\exp(B_{\delta}^{\h})H_0$ is a nghd of $H_0$.

Given a sequence $H_i \in \exp(\h)$ converging to $H_0$, for $n$ large enough $H_n \in \exp(B_{\delta}^{\h})H_0$. Then $H_0$ is in invariant manifold $\exp(\h)$. Hence  $\exp(\h)$ is
closed and hence compact.

Let $y \in \f$, then there exists a $h_0 \in \h$ such that
$ \exp(h_0) y \exp(-h_0) \in \a$. We maximize the function
$ \langle a_r, \exp(h) y \exp(h) \rangle $, over the compact group $\exp(\h)$, for regular element $a_r \in \a$ and $\langle .,.\rangle$ is the killing form. At the maxima, we have at $t=0$, $\frac{d}{dt} \langle a_r, \exp(h_1 t) (\exp(h_0) y \exp(-h_0)) \exp(-h_1 t) \rangle = 0$.
$$ \langle a_r, [h_1 \exp(h_0) y \exp(-h_0)] \rangle = - \langle h_1, [a_r \exp(h_0) y \exp(-h_0)] \rangle, $$

if $\exp(h_0) y \exp(-h_0) \neq \a$, then $[a_r,\  \exp(h_0) y \exp(-h_0)] \in \k$. The bracket
$[a_r,\  \exp(h_0) y \exp(-h_0)]$ is $Ad_{A^2}$ invariant and hence belong to $\h$. We can choose
$h_1$ so that gradient is not zero. Hence $\exp(h_0) y \exp(-h_0) \in \a$. For $z \in \p$ such that $z \in \f^{\perp}$, we have $\exp(h_0) z \exp(-h_0) \in \a^{\perp}$.
$$ \langle \a, \exp(h_0) z \exp(-h_0) \rangle  = \langle \exp(-h_0) \a \exp(h_0), z  \rangle = 0, $$
as $\exp(-h_0) \a \exp(h_0)$ is $Ad_{A^2}$ invariant, hence
$\exp(-h_0) \a \exp(h_0) \in \f$. In above, we worked with killing form. For $\g = su(n)$, we may use standard inner product.}\end{remark}

\begin{remark}{\bf Kostant Convexity}\label{rem:convexity}{ \rm  \cite{Kostant} Given the decomposition $\g = \p \oplus \k$, let $\a \subset \p$ and $X \in \a$,. Let $\W_i \in \exp(\k)$ such that $\W_i X \W_i \in \a$ are distinct, Weyl points. Then projection (w.r.t killing form) of $Ad_K(X)$ on $\a$ lies in convex hull of these Weyl points. The $\C$ be the convex hull and let projection $P(Ad_K(X))$ lie outside this Hull. Then there is a separating hyperplane $a$, such that $\langle Ad_K(X), a \rangle <  \langle \C, a \rangle $. W.L.O.G we can take $a$ to be a regular element. We minimize
$\langle Ad_K(X), a \rangle$, with choice of $K$ and find that minimum happens when $[Ad_K(X), a] = 0$, i.e. $Ad_K(X)$ is a Weyl point. Hence $P(Ad_K(X)) \in \sum_i \alpha_i \W_i X \W_i^{-1}$, for $\alpha_i > 0$ and $\sum_i \alpha_i = 1$. The result is true with a projection w.r.t inner product that satisfies $\langle x, [y, z] \rangle = \langle [x, y], z] \rangle $, like standard inner product on $\g = su(n)$.} \end{remark}

We now give an elementary proof (using eigenvalues, eigenvectors) of the special case of KAK decomposition for the group $G = SU(2n)$ with a closed subgroup $K = SU(n) \times SU(n) \times U(1)$, of block diagonal special unitaries, such that the respective lie algebras $\g = su(2n)$, traceless skew hermitians, and $\k = su(n) \oplus su(n) \oplus u(1)$, block diagonal, traceless skew hermitians, have the Cartan decomposition $\g = \p \oplus \k$ where $\p = \left [ \begin{array}{cc} 0 & Z \\ -Z' & 0 \end{array} \right ]$. The associated cartan subalgebra $\a = \left [ \begin{array}{cc} 0 & \lambda \\ -\lambda & 0 \end{array} \right ]$ where $\lambda = \left[ \begin{array}{cccc} \lambda_1 & 0 & \hdots & 0 \\  0 & \lambda_2 & \hdots & 0 \\  \vdots & \vdots & \ddots & \vdots \\ 0 & 0 & \hdots & \lambda_n \end{array} \right ]$ and $|\lambda_i| \neq |\lambda_j| \neq 0$ is a regular element of $\a$.

\begin{theorem}\label{th:SU(2n)decompose}{\rm Let $U \in SU(2n)$, then
$$ U = \left [ \begin{array}{cc} K_1 & 0  \\  0 & K_2 \end{array} \right ] \exp(\left[ \begin{array}{cc} 0 & \lambda  \\  -\lambda & 0 \end{array} \right ]) \left[ \begin{array}{cc} K_3 & 0  \\  0 & K_4 \end{array} \right ], $$where $\left [ \begin{array}{cc} K_1 & 0  \\  0 & K_2 \end{array} \right ], \left [ \begin{array}{cc} K_3 & 0  \\  0 & K_4 \end{array} \right ]  \in SU(n)\times SU(n)\times U(1)$ (Block diagonal special unitary matrices) and $$ \left[ \begin{array}{cc} 0 & \lambda  \\  -\lambda & 0 \end{array} \right ] =  \left[ \begin{array}{cccccc} 0 & \hdots & 0 & \lambda_1 & \hdots & 0  \\   \vdots  & \ddots & \vdots & \vdots & \ddots & \vdots \\  0 & \hdots & 0 & \hdots & 0 & \lambda_n \\  -\lambda_1 & \hdots & 0 & 0 & \hdots & 0 \\  \vdots  & \ddots & \vdots & \vdots & \ddots & \vdots \\ 0 & \hdots & -\lambda_n & 0 & \hdots & 0 \end{array} \right ] $$ }\end{theorem}

{\bf Proof: }Let Block diagonal $$ S = \left[ \begin{array}{cc} \Iden & 0  \\  0 & -\Iden \end{array} \right ] $$

Then for $\cos(\lambda) =  \left[ \begin{array}{cccc} \cos \lambda_1 & 0 & \hdots & 0 \\  0 & \cos \lambda_2 & \hdots & 0 \\  \vdots & \vdots & \ddots & \vdots \\ 0 & 0 & \hdots & \cos \lambda_n \end{array} \right ]$ 
$$ S \left[ \begin{array}{cc} \cos(\lambda) & -\sin(\lambda)  \\  \sin(\lambda) & \cos(\lambda) \end{array} \right ]S = \left[ \begin{array}{cc} \cos(\lambda) & \sin(\lambda)  \\  -\sin(\lambda) & \cos(\lambda) \end{array} \right ] $$

$USU'S \in SU(2n)$, then let $x$ be an eigenvector of $USU'S$. Then

$$ USU'S x = \exp(j \theta) x. $$ Taking inverse

$$ USU'S (Sx) = \exp(-j \theta) Sx. $$

Let $\Sigma_1$, be perpendicular eigenvectors corresponding to eigenvalues $\exp(-j \theta)$.

Then $S \Sigma_1$ are perpendicular eigenvectors corresponding to eigenvalues $\exp(j \theta)$.

This says that eigenvalues $\exp(j \theta)$ and $\exp(-j \theta)$ have same multiplicities.

This leaves us with eigenvalues $1$ and $-1$. Given the eigenvector

$ z = \left[ \begin{array}{c} x \\  y \end{array} \right ] $, with eigenvalue $1$, $ S z = \left[ \begin{array}{c} x \\  -y \end{array} \right ]$ is an eigenvector with eigenvalue $1$. This says eigenvectors of $1$ and $-1$ are of the form $ \left[ \begin{array}{c} x \\  0 \end{array} \right ] $  and $ \left[ \begin{array}{c} 0 \\  y \end{array} \right ] $. This allows to form orthonormal pairs $ \left[ \begin{array}{c} x \\  y \end{array} \right ] $ and $ \left[ \begin{array}{c} x \\  - y \end{array} \right ]$. After pairing, let $x_{\pm}$ and $y_{\pm}$ be surplus eigenvectors with $\pm 1$ eigenvalues in which $y$ and $x$ parts are zero respectively. Then, dimension of independent $x_{+}$ and $y_{-}$ is the same. The eigenvectors can be organized in columns as follows
$$ P = \left[ \begin{array}{cccc} z_{+} \ x_{+} \ z_{-} \ y_{-}  \end{array} \right ], $$ where $z_{+}$ and $z_{-}$ are eigenvectors corresponding to $ \left[ \begin{array}{c} x \\  y \end{array} \right ] $, and $ \left[ \begin{array}{c} x \\  -y \end{array} \right ]$ respectively.

Let $I= \frac{1}{\sqrt{2}} \Iden_{k \times k} $ and $I_1 = \Iden_{m \times m}$ and $I_0 = \Ze_{m \times m}$.

$$ S_1 = \left[ \begin{array}{cc} I_A & -I_B \\ I_B & I_A \end{array} \right ]. $$

$$ I_A = \left[ \begin{array}{cc} I & 0 \\ 0 & I_1 \end{array} \right ], \ \  I_B = \left[ \begin{array}{cc} I & 0 \\ 0 & I_0 \end{array} \right ]. $$

 Then $$ USU'S  = P S_1 S_1' \left[ \begin{array}{cccc} I_{\theta} & 0 & 0 & 0 \\ 0 & I_1 & 0 & 0
  \\ 0 & 0 & I_{-\theta} & 0 \\ 0 & 0 & 0 & -I_1 \end{array} \right ]S_1 S_1' P',  $$ where,

$I_{\theta} = \left[ \begin{array}{cccc} \exp(i \theta_1) & 0 & 0 & 0 \\ 0 & \exp(i \theta_2) & 0 & 0
  \\ 0 & 0 & \ddots & 0 \\ 0 & 0 & 0 & \exp(i \theta_k) \end{array} \right ]$.

Then $$ P S_1 = \left[ \begin{array}{cc} K_1 & 0 \\ 0 & K_2 \end{array} \right ] = K .$$

Then $$ S_1' \left[ \begin{array}{cccc} I_{\theta} & 0 & 0 & 0 \\ 0 & I_1 & 0 & 0
  \\ 0 & 0 & I_{-\theta} & 0 \\ 0 & 0 & 0 & -I_1 \end{array} \right ]S_1 = \left[ \begin{array}{cccc} \cos_\theta & 0 & -i \sin_\theta & 0  \\ 0 & I_1 & 0 & 0 \\ -i \sin_\theta & 0 & \cos_\theta & 0 \\ 0 & 0 & 0 & -I_1    \end{array} \right ]= A^2. $$

Let $$ U = \left[ \begin{array}{cccc} K_1 & 0 \\ 0 & K_2 \end{array} \right ] \left[ \begin{array}{cccc} \cos_{\frac{\theta}{2}} & 0 & -i \sin_{\frac{\theta}{2}} & 0 \\ 0 & I_1  & 0 & 0 \\ -i \sin_{\frac{\theta}{2}} & 0 & \cos_{\frac{\theta}{2}} & 0 \\ 0 & 0 & 0 & i I_1 \end{array} \right ] V = K A V. $$

Then $$ VSV' = A' K'U S U' K A = A' K' U S U' S S K A = A' K' K A^2 K' K A'' S = W S. $$

where,

$$ A'' = \left[ \begin{array}{cccc} \cos_{\frac{\theta}{2}} & 0 & i \sin_{\frac{\theta}{2}} & 0 \\ 0 & I_1  & 0 & 0 \\ i \sin_{\frac{\theta}{2}} & 0 & \cos_{\frac{\theta}{2}} & 0 \\ 0 & 0 & 0 & i I_1 \end{array} \right ]. $$

This gives $SVS = W V$,where $ V = \left[ \begin{array}{cc} U_1 & U_2 \\ U_3 & U_4 \end{array} \right ]$.

$W = \left[ \begin{array}{ccc} \Iden_{n \times n} & 0 & 0 \\ 0 & I & 0 \\ 0 & 0 & -I_1 \end{array} \right ] $.

This gives $U_2 = 0$. Since $V$ is unitary, gives $U_3 = 0$. Therefore $U=$

$$\left[ \begin{array}{cccc} K_1 & 0 \\ 0 & K_2 \end{array} \right ] \left[ \begin{array}{cccc} \cos_{\frac{\theta}{2}} & 0 & -i \sin_{\frac{\theta}{2}} & 0 \\ 0 & I_1  & 0 & 0 \\ -i \sin_{\frac{\theta}{2}} & 0 & \cos_{\frac{\theta}{2}} & 0 \\ 0 & 0 & 0 & i I_1 \end{array} \right ] \left[ \begin{array}{cccc} U_1 & 0 \\ 0 & U_4 \end{array} \right ] =
 \left[ \begin{array}{cccc} K_1 & 0 \\ 0 & K_2 \end{array} \right ] \left[ \begin{array}{cccc} \cos_{\frac{\theta}{2}} & 0 & \sin_{\frac{\theta}{2}} & 0 \\ 0 & I_1  & 0 & 0 \\ -\sin_{\frac{\theta}{2}} & 0 & \cos_{\frac{\theta}{2}} & 0 \\ 0 & 0 & 0 & I_1 \end{array} \right ] \left[ \begin{array}{cccc} K_3 & 0 \\ 0 & K_4 \end{array} \right ] , $$ where $ \left[ \begin{array}{cccc} K_1 & 0 \\ 0 & K_2 \end{array} \right ]$ and  $\left[ \begin{array}{cccc} K_3 & 0 \\ 0 & K_4 \end{array} \right ]$ are block diagonal special unitary matrices. {\bf q.e.d}

\n We now give a proof of the reachable set in Eq. (\ref{eq:reachableintro}), for the $\frac{SU(2n)}{SU(n) \times SU(n) \times U(1)}$ problem.


\begin{theorem}\label{th:model2}{\rm   Let $P(t) \in SU(2n)$ be a solution to the differential equation $$ \dot{P} = Ad_{K(t)}(X_d) P, $$ where $K(t) = \left [ \begin{array}{cc} K_1  & 0 \\ 0 & K_2 \end{array} \right ] \in SU(n) \times SU(n) \times U(1)$, block diagonal special unitary matrices. $$ X_d =  \left[ \begin{array}{cc} 0 & \lambda  \\  -\lambda & 0 \end{array} \right ] =  \left[ \begin{array}{cccccc} 0 & \hdots & 0 & \lambda_1 & \hdots & 0  \\   \vdots  & \ddots & \vdots & \vdots & \ddots & \vdots \\  0 & \hdots & 0 & \hdots & 0 & \lambda_n \\  -\lambda_1 & \hdots & 0 & 0 & \hdots & 0 \\  \vdots  & \ddots & \vdots & \vdots & \ddots & \vdots \\ 0 & \hdots & -\lambda_n & 0 & \hdots & 0 \end{array} \right ]. $$ The elements of reachable set at time $T$, takes the form
 $$ P(T) = \left [ \begin{array}{cc} \Theta_1  & 0 \\ 0 & \Theta_2 \end{array} \right ]\exp (T \sum_k \alpha_k \W_k (\left [ \begin{array}{cc} 0 & \lambda \\ -\lambda & 0 \end{array} \right ])\W_k') \left [ \begin{array}{cc} \Theta_3  & 0 \\ 0 & \Theta_4 \end{array} \right ], $$ where $ \left [ \begin{array}{cc} \Theta_1  & 0 \\ 0 & \Theta_2 \end{array} \right ],   \left [ \begin{array}{cc} \Theta_3  & 0 \\ 0 & \Theta_4 \end{array} \right ], \W_k  \in SU(n) \times SU(n) \times U(1)$. $\W_k$ induce permutations and sign changes on $\lambda$.}\end{theorem}

From theorem \ref{th:SU(2n)decompose}, for every time $t$, $P(t)$ has the form, $$ P =  \underbrace{\left [ \begin{array}{cc} U_1  & 0 \\ 0 & U_2 \end{array} \right ]}_{K_1}  \left[ \begin{array}{cccccc} \cos \phi_1 & \dots & 0 & \sin \phi & \hdots & 0 \\  \vdots & \ddots & \vdots & \vdots & \ddots & \vdots \\  0 & 0 & \cos \phi_n & 0 & 0 & \sin \phi_n   \\  -\sin \phi_1 & 0 & 0 & \cos \phi_1  & \dots & 0 \\  \vdots & \ddots & \vdots & \vdots & \ddots  & \vdots \\  0
 & 0 & -\sin \phi_n & 0 & 0 & \cos \phi_n \end{array} \right ]  \underbrace{\left [ \begin{array}{cc} U_3  & 0 \\ 0 & U_4 \end{array} \right ]}_{K_2}, $$

$$ P =  K_1 \exp \left[ \begin{array}{cccccc} 0 & \dots & 0 & \phi_1 & \hdots & 0 \\  \vdots & \ddots & \vdots & \vdots & \ddots & \vdots \\  0 & 0 & 0 & 0 & 0 & \phi_n   \\  -\phi_1 & 0 & 0 & 0 & \dots & 0 \\  \vdots & \ddots & \vdots & \vdots & \ddots  & \vdots \\  0
 & 0 & -\phi_n & 0 & 0 & 0 \end{array} \right ] K_2 , $$

As in theorem \ref{th:model1}, we consider evaluating the product  $\Pi \exp(Ad_{K_i}(X_d)\Delta )$. We approximate the evolution
$\exp(Ad_K(X_d)\Delta ) P(t)$ by $Q(t + \Delta)$, where we define $Q(t + \Delta)$ as

\begin{equation}
\label{eq:Qdegenerate.2}
 Q(t + \Delta) = \exp(\Omega_1 \Delta) K_1 H_1 \exp(a \Delta) A H_2 \exp(\Omega_2 \Delta) K_2
\end{equation}

where $H_1, H_2 \in \exp(\h)$, the stabilizer group, as discussed in remark \ref{rem:stabilizer}, such that $H_1 A H_2 = A$, by choosing $H_2^{-1} = A^{-1}H_1A$.

The above expression can be written as

$$ Q(t + \Delta) = \exp(\Omega_1 \Delta) \exp(K_1 H_1 a H_1'K_1' \Delta) \exp(K_1 A \Omega_2 A' K_1' \Delta) P(t). $$

Where $\Omega_1$, $H_1, a, \Omega_2$, have the same meaning as in theorem \ref{th:model1} in context of the present decomposition $\g = \p \oplus \k$  and are chosen such that

$$ (\Omega_1 + K_1 H_1 a H_1' K_1' + K_1 A \Omega_2 A' K_1') = Ad_K(X_d). $$

$$ (\Omega_1' + H_1 a H_1' + A \Omega_2 A') = Ad_{\bar K}(X_d), $$ where $\bar{K} = K_1^{-1}K$ and

\begin{equation}
P(t + \Delta) = \exp( (\Omega_1 + K_1 H_1 a H_1' K_1' + K_1 A \Omega_2 A' K_1') \Delta) P(t) \end{equation}

$$ P(t + \Delta) - Q(t + \Delta) = o(\Delta^2) P(t) $$

$$ Q(t + \Delta) = (I + o(\Delta^2))P(t + \Delta). $$

Let

$$ \left [ \begin{array}{cc} A  & B \\ C & D \end{array} \right ]^{\ast} =  \left [ \begin{array}{cc} I & 0 \\ 0 & -I \end{array} \right ] \left [ \begin{array}{cc} A  & B \\ C & D \end{array} \right ]' \left [ \begin{array}{cc} I  & 0 \\ 0 & -I \end{array} \right ], $$ where we have $n \times n$ subblocks.

$$ PP^\ast =  \left [ \begin{array}{cc} K_1  & 0 \\ 0 & K_2 \end{array} \right ]  \left[ \begin{array}{cccccc} \cos 2\phi_1 & \dots & 0 & \sin 2\phi & \hdots & 0 \\  \vdots & \ddots & \vdots & \vdots & \ddots & \vdots \\  0 & 0 & \cos 2\phi_n & 0 & 0 & \sin 2\phi_n   \\  -\sin 2\phi_1 & 0 & 0 & \cos 2\phi_1  & \dots & 0 \\  \vdots & \ddots & \vdots & \vdots & \ddots  & \vdots \\  0
 & 0 & -\sin 2\phi_n & 0 & 0 & \cos 2\phi_n \end{array} \right ]  \left [ \begin{array}{cc} K_1'  & 0 \\ 0 & K_2' \end{array} \right ],$$

let $S = \exp(-i \frac{\pi}{2} \sigma_x \otimes \Iden_{n \times n})$

$$ PP^\ast =  \left [ \begin{array}{cc} K_1  & 0 \\ 0 & K_2 \end{array} \right ]  S' \left[ \begin{array}{cccccc} \exp(i 2\phi_1) & \dots & 0 & 0 & \hdots & 0 \\  \vdots & \ddots & \vdots & \vdots & \ddots & \vdots \\  0 & 0 & \exp(i 2\phi_n) & 0 & 0 & 0   \\  0 & 0 & 0 & \exp(-i 2\phi_1)  & \dots & 0 \\  \vdots & \ddots & \vdots & \vdots & \ddots  & \vdots \\  0
 & 0 & 0 & 0 & 0 & \exp(-i 2\phi_n) \end{array} \right ]  S \left [ \begin{array}{cc} K_1'  & 0 \\ 0 & K_2' \end{array} \right ],$$

The above expression can be written as

$$ Q(t + \Delta)Q(t + \Delta)^{\ast} = (I + o(\Delta^2))P(t + \Delta)P^{\ast}(t + \Delta) (I + o(\Delta^2)) = P(t+ \Delta)P^{\ast}(t + \Delta) [ I + o(\Delta^2)]. $$

$Q(t+\Delta)Q(t + \Delta)^{\ast} = \underbrace{\left [ \begin{array}{cc} \tilde{K}_1  & 0 \\ 0 & \tilde{K}_2 \end{array} \right ]  S^{\dagger}}_{U} \Sigma \underbrace{S \left [ \begin{array}{cc} \tilde{K}_1  & 0 \\ 0 & \tilde{K}_2 \end{array} \right ]'}_{U'}$, where,
$$\Sigma = \left[ \begin{array}{cccccc} \exp i2(\phi_1 + a_1 \Delta) & \dots & 0 & 0 & \hdots & 0 \\  \vdots & \ddots & \vdots & \vdots & \ddots & \vdots \\  0 & 0 & \exp i 2(\phi_n + a_n \Delta) & 0 & 0 & 0  \\  0 & 0 & 0 & \exp -i 2(\phi_1 + a_1 \Delta)  & \dots & 0 \\  \vdots & \ddots & \vdots & \vdots & \ddots  & \vdots \\  0
 & 0 & 0 & 0 & 0 & \exp -i 2(\phi_n + a_n \Delta) \end{array} \right ].$$

 $$ P(t + \Delta)P^{\ast}(t+\Delta) =  V \left[ \begin{array}{cccccc} \exp i \mu_1  & \dots & 0 & 0 & \hdots & 0 \\  \vdots & \ddots & \vdots & \vdots & \ddots & \vdots \\  0 & 0 & \exp i \mu_n & 0 & 0 & 0  \\  0 & 0 & 0 & \exp -i \mu_1  & \dots & 0 \\  \vdots & \ddots & \vdots & \vdots & \ddots  & \vdots \\  0
 & 0 & 0 & 0 & 0 & \exp -i \mu_n \end{array} \right ]V^{\dagger} ,$$

By choosing $\Delta$ small, we can as in theorem \ref{th:model1}, bound $|o(\Delta^2)| < c \Delta $ (see Eq. \ref{eq:nghdbnd}). If $\tilde{a} = \min_{a_i \neq \pm a_j}(a_i \pm a_j)$. For small $\Delta$, the minimum spacing between non-degenerate eigenvalues of $QQ^{\ast} (t + \Delta)$ is $\tilde{a} \Delta$.  When $c \leq \frac{\tilde{a}}{6}$, $c \Delta \leq \frac{\tilde{a} \Delta}{6} \leq 2 \sin(\frac{\tilde{a} \Delta}{6})$.

If eigenvalue
$\exp(-i \lambda_j)$ in $\lambda(QQ^{\ast})= \left [ \begin{array}{c} \exp(-i \lambda) \\ \exp(i \lambda) \end{array} \right ]$ ($\exp(-i \lambda)$ is an eigenvalue set)), is $m+n$ degenerate ($m$ coming from top, remaining from bottom), then its $\frac{\tilde{a} \Delta}{3}$, nghd, has precisely $m+n$ eigenvalues of $\lambda(PP^{\ast})$. This follows from Schur convexity, as otherwise

$$ o (\Delta^2) = |QQ^{\ast} -PP^{\ast}| = \sum_k \alpha_k |\phi - P_k(\mu)| > 2 \sin(\frac{\tilde{a} \Delta}{6}), $$ where 
$\phi = \lambda(QQ^{\ast}), \mu = \lambda(PP^{\ast})$, $\sum_k \alpha_k = 1$ and $P_k$ are permulations. Of these $m+n$ eigenvalues of $\lambda(PP^{\ast})$, $m$ can be assigned to  pocket of $\exp(-i \lambda_j)$ and remaining $n$ to conjugate pocket  $\exp(i \lambda_j)$ ($m$ in top, $n$ in bottom). We have shown an eigenvalue set of $PP^{\ast}$ which is in ngd of eigenvalue set of $QQ^\ast$.

We can again set up a chain of overlapping nghds such that evolution of eigenvalues
can be written as

\begin{eqnarray}
\lambda(Q_{i+}Q_{i+}^{\ast}) &=& \exp(2 a_i^{+} \Delta_i^+)\ \lambda(P_{i}P_{i}^{\ast})  \\
\lambda(P_{i, i+1}P_{i, i+1}^{\ast}) &=& \lambda(Q_{i+}Q^{+}_{i+}) + o((\Delta_i^+)^2) \\
\lambda(Q_{(i+1)-}Q_{(i+1)-}^{\ast}) &=& \lambda(P_{i, i+1}P_{i, i+1}^{\ast}) + o((\Delta^{-}_{i+1})^2) \\
\exp(-2 a_{i+1}^{-} \Delta_{i+1}^{-})\ \lambda(P_{i+1}P_{i+1}^{\ast}) &=& \lambda(Q_{(i+1)-}Q^{\ast}_{(i+1)-})
\end{eqnarray}where in Eq. (\ref{eq:Qdegenerate.2}), $a$ has the form, $a = \left[ \begin{array}{cc} 0 & a_i^{+} \\ -a_i^{+} & 0 \end{array} \right]$ and $\lambda(QQ^{\ast})$ only denotes one (say top one) of the conjugate eigenvalue set. Adding the above equations,

\begin{equation}
\lambda(P_{i+1}P_{i+1}^{\ast}) = \exp (o(\Delta^2))\ \exp(2 ( a_i^+ \Delta_i^+ + a_{i+1}^{-}\Delta_{i+1}^{-}))\ \lambda(P_i P_i^{\ast}),
\end{equation} where $o(\Delta^2)$ is diagonal.

where $|\exp(i \alpha) -1| = 2 \sin\frac{|\alpha|}{2} > \frac{|\alpha|}{2}$. Therefore, $|\exp(i \alpha) - 1| \leq \frac{\theta}{2}$ implies $|\alpha| \leq \theta$.

\begin{equation}
\lambda(P_nP_n^{\ast}) =  \exp(\underbrace{\sum o(\Delta^2)}_{\leq \epsilon T})\ \exp(2 \sum_i a_i^{+} \Delta_i^{+} + a_{i+1}^{-}\Delta_{i+1}^{-})\ \lambda(P_1P_1^{\ast}).
\end{equation}

\begin{equation}
\lambda(P_nP_n^{\ast}) = \exp(\underbrace{\sum o(\Delta^2)}_{\leq \epsilon T})\ \exp(2 T \sum_k \alpha_k P_k(\lambda))\ \lambda(P_1P_1^{\ast}) =  \exp(\underbrace{\sum o(\Delta^2)}_{\leq \epsilon T})\ \exp(2 \mu T)\ \lambda(P_1P_1^{\ast}),
\end{equation}where $P_k$ are $2^n n!$ permutations alongwith sign changes. By Kostant convexity in remark \ref{rem:convexity},  $\mu \prec \lambda$, i.e. $\mu$ lies in convex hull of permutation and sign changes of $\lambda$. Since $P_1 = I$,

\begin{equation}
P_n = K_1 \exp \left[ \begin{array}{cccccc} 0 & \dots & 0 & \mu_1T +  m_1 \pi & \hdots & 0 \\  \vdots & \ddots & \vdots & \vdots & \ddots & \vdots \\  0 & 0 & 0 & 0 & 0 & \mu_nT + m_n \pi   \\  -( \mu_1T +  m_1 \pi)  & 0 & 0 & 0 & \dots & 0 \\  \vdots & \ddots & \vdots & \vdots & \ddots  & \vdots \\  0
 & 0 & -( \mu_nT +  m_n \pi) & 0 & 0 & 0 \end{array} \right ] \exp(o(\Delta^2)) K_2.
\end{equation}

$m \pi$ can be absorbed in $K_1$. By letting $\Delta^2$, go to zero, we find
$$ P_n \in K_1 \exp (T \sum_k \alpha_k \W_k (\left [ \begin{array}{cc} 0 & \lambda \\ -\lambda & 0 \end{array} \right ])\W_k') K_2. $$

where $\W_k$ are Weyl elements that induce permutations and sign changes of $\lambda$.

\begin{corollary}\label{cor:en}{\rm
\begin{equation}
\dot{U} = ( X_d + \sum_j u_j(t) X_j ) U,  \ \ U(0) = \Iden.
\end{equation}
Here $X_d = \left[ \begin{array}{cc} 0 & \lambda \\ -\lambda & 0 \end{array} \right ]$ and  $\{X_j\}_{LA} = \left [ \begin{array}{cc} A  & 0 \\ 0 & B \end{array} \right ]$, space of block diagonal skew Hermitian matrices. The elements of reachable set at time $T$, takes the form
$ U(T) \in$  $$ S = K_1 \exp (T \sum_k \alpha_k \W_k (\left [ \begin{array}{cc} 0 & \lambda \\ -\lambda & 0 \end{array} \right ])\W_k') K_2, $$ where $\W_k$ are Weyl elements that induce permutations and sign changes of $\lambda$ and $K_1, K_2, \W_k \in SU(n)\times SU(n) \times U(1)$ (Block diagonal matrices in $SU(2n)$ ). $S$ belongs to closure of reachable set. }\end{corollary}

\begin{example}{\rm The problem in this section models the evolution of electron-nuclear spin system in EPR \cite{Zeier}. Let $S$ and $I$ be spin operators for electron and nuclear spin respectively. The evolution of coupled spin dynamics is in $G = SU(4)$. Its algebra $$ \g = -i \{ I_{\alpha}, S_\beta, I_{\alpha}S_{\beta} \} = \underbrace{-i \{I_x, I_y, I_x S_\alpha, I_yS_{\beta}\}}_{\p} \oplus -i \underbrace{\{I_z, S_{\alpha}, I_{z}S_{\beta}\}}_{\k}. $$ $\k$ are fast spin operators in the dynamics and involve rotations derived from electron spin rotations, hyperfine coupling and Larmor precession of nuclear spin. $\p$ represent operators that rotate nuclear spin with rf-pulses and are slow part of dynamics. The control subgroup is $K = \exp(\k) = SU(2) \times SU(2) \times U(1)$. The Cartan subalgebra is $\a = -i \{I_x, I_xS_z \}$. 
Given $X_d \in \a$ such that $X_d = -i(\alpha I_x  + \beta 2 I_x S_z) = -i [ \underbrace{(\alpha+\beta)}_{2a} I_x(\frac{\Iden}{2} + S_z) + \underbrace{(\alpha -\beta)}_{2b} I_x(\frac{\Iden}{2} -S_z)] $. 
$$ X_d = -i \left [ \begin{array}{cc} 0 & \Lambda \\ \Lambda & 0 \end{array} \right ] ; \ \ \Lambda = \left [ \begin{array}{cc} a & 0 \\ 0 & b \end{array} \right ]. $$ There are eight Weyl Points, $(a, b)$, and its permutations with sign changes. We ask for a reachable set described by evolution of the kind $$ U = K_0 \prod_i \exp(X_d \Delta t_i ) K_i, \ \ \sum \Delta t_i = T, \ \ K_i, K_0 \in K $$

The reachable set is described in above corollary \ref{cor:en}. }\end{example}

We now present results in section \ref{sec:second} and \ref{sec:third} in a more general setting.

\section{Time Optimal control for $G/K$ problem}
\label{sec:fourth}


\begin{theorem}\label{th:generalmodel}{\rm Given a compact Lie group $G$ and Lie algebra $\g$. Consider the Cartan decomposition of a real semisimple Lie algebra $\g = \p \oplus \k$. Given the control system
$$ \dot{X} = Ad_{K(t)}(X_d) X, \ P(0) = \Iden $$ where $X_d \in \a$,
    the cartan subalgebra $\a \in \p$  and $K(t) \in \exp{\k}$, a closed subgroup of $G$. The
    end point $$P(T) = K_1 \exp(T \sum_i \alpha_i \W_i(X_d) ) K_2, $$
    where $K_1, K_2 \in \exp(\k)$ and $\W_i(X_d) \in \a$ are Weyl points, $\alpha_i > 0$ and $\sum_i {\alpha_i} = 1$.} \end{theorem}

\no {\bf Proof:} As in proof of Theorem \ref{th:model1} and \ref{th:model2}, we define $$ P(t + \Delta) =  \exp(Ad_K(X_d) \Delta ) P(t) = \exp(Ad_K(X_d) \Delta ) K_1 \exp(a) K_2 $$ and show that

\begin{equation}
\label{eq:iterative.a.increment} 
\exp(Ad_K(X_d) \Delta)K_1 A K_2  = K_a \exp(a_0 \Delta + C \Delta^2) A K_b = K_a \exp(a + a_0 \Delta + C \Delta^2) K_b,
\end{equation}

where for $\bar{K} = K^{-1}K$,

$$ Ad_{\bar{K}}(X_d) = \underbrace{P(Ad_{\bar{K}}(X_d))}_{a_0} + Ad_{\bar{K}}(X_d)^{\perp}.$$

where $P$ is projection w.r.t killing form and $a_0 \in \f$, the centralizer in $\p$ as defined in remark \ref{rem:stabilizer}, $C \Delta^2 \in \f$ is a second order term that can be made small by choosing $\Delta$ and $K_a, K_b \in \exp(\k)$.

To show Eq. \ref{eq:iterative.a.increment}, we show there exists $K_1'', K_2'' \in K$ such that 

\begin{equation}
\label{eq:iterative.a.increment1} 
\underbrace{\exp(k_1'')}_{K_1''} \exp(Ad_{\bar K}(X_d) \Delta) \underbrace{\exp(A k_2'' A^{-1})}_{K_2''} = \exp(a_0 \Delta + C \Delta^2),
\end{equation} where $K_1''$ and $K_2''$ are constructed by a iterative procedure as described in the proof below.

Given $X$ and $Y$ as $N \times N$ matrices, considered elements of a matrix Lie algebra $\g$, we have,
\begin{equation}
log(e^X e^Y) - (X + Y) = \sum_{n>0} \frac{(-1)^{n-1}}{n} \sum_{1\leq i \leq n}  \frac{[X^{r_1}Y^{s_1}\dots X^{r_n}Y^{s_n}]}{\sum_{i=1}^n (r_i + s_i) r_1! s_1! \dots r_n! s_n !},
\end{equation}where $r_i + s_i > 0$.

We bound the largest element (absolute value) of $log(e^X e^Y) - (X + Y)$, denoted as $|log(e^X e^Y) - (X + Y)|_0$, given
$|X|_0 < \Delta$ and $|Y|_0 < b_0 \Delta^k$ , where $k \geq 1$, $\Delta < 1$, $b_0 \Delta < 1 $.

\begin{eqnarray}
\label{eq:bound}
|log(e^X e^Y) - (X + Y)|_0 &\leq&  \sum_{n=1} N b_0 e \Delta^{k+1} + \sum_{n>1} \frac{1}{n} \frac{(2Ne^2)^n b_0 \Delta^{n+k-1}}{n} \\
&\leq &  N b_0 e \Delta^{k+1} + (Ne^2)^2 b_0 \Delta^{k+1} ( 1 + 2Ne^2\Delta + \dots ) \\
&\leq &  N b_0 e \Delta^{k+1} + \frac{(Ne^2)^2 b_0 \Delta^{k+1}}{1 - 2Ne^2\Delta} \leq \tilde{M} b_0 \Delta^{k+1}
\end{eqnarray}where $2 N \Delta < 1$ and $\tilde{M} \Delta < 1$.

Given decomposition of $\g = \p \oplus \k $, $\p \perp \k $ with respect to the negative definite killing form
$B(X,Y) = tr(ad_X ad_Y)$. Furthermore there is decomposition of $\p = \a \oplus \a^{\perp}$.

Given $$ U_0 = \exp(a_0 \Delta + b_0 \Delta + c_0 \Delta), $$ where
$a_0 \in \a$, $b_0 \in \a^{\perp}$ and $c_0 \in \k$, such that $|a_0|_0 + |b_0|_0 + |c_0|_0 < 1$, which we just abbreviate as $a_0 + b_0 + c_0 < 1$
(we follow this convention below)

We describe an iterative procedure

\begin{equation}
U_n = \Pi_{k=1}^n \exp(-c_k \Delta ) \ U_0 \ \Pi_{k=0}^n \exp(-b_{k} \Delta),
\end{equation}

where $c_k \in \k$ and $b_k \in \a^{\perp}$, such that the limit

\begin{equation}
n \rightarrow \infty \ \ U_n = \exp(a_0 \Delta + C \Delta^2),
\end{equation}where $a_0, C \in \a$.

\begin{eqnarray*}
U_1 &=& \exp(-c_0 \Delta)\exp(a_0 \Delta + b_0 \Delta + c_0 \Delta) \exp(-b_0 \Delta) \\
    &=& \exp(a_0 \Delta + b_0 \Delta + c_0' \Delta^2) \exp(-b_0 \Delta) \\
    &=& \exp(a_0 \Delta + b_0' \Delta^2 + c_0' \Delta^2) \\
    &=& \exp( (a_1 + b_1 + c_1 )\Delta )
\end{eqnarray*}
Note $b_0'$ and $c_0'$ are elements of $\g$ and need not be contained in $\a^{\perp}$ and $\k$.

Where, using bound in $c_0' \leq \tilde{M}c_0$, which gives $a_0 + b_0 + c_0' \Delta \leq a_0 + b_0 + c_0$. Using the bound again,
we obtain, $b_0' \leq \tilde{M} b_0 $. We can decompose, $(b_0' + c_0')\Delta$, into subspaces $a_0'' + b_1 + c_1$, where
$a_0'' \leq M (b_0' + c_0')\Delta$, $b_1 \leq M(b_0' + c_0')\Delta$ and $c_1 \leq M(b_0' + c_0')\Delta$, where $-B(X, X) \leq \lambda_{max} |X|^2$, where $|X|$ is Frobenius norm and $-B(X, X) \geq \lambda_{min} |X|^2$. Let $M = \frac{N \lambda_{max}}{\lambda_{min}}$.

This gives, $a_0'' \leq M(b_0' + c_0')\Delta$, $b_1 \leq M(b_0' + c_0')\Delta$ and $c_1 \leq M(b_0' + c_0')\Delta$. This gives

\begin{eqnarray*}
a_1 &\leq& a_0 + \tilde{M}M (b_0 + c_0)\Delta \\
b_1 &\leq& \tilde{M}M(b_0 + c_0) \Delta \\
c_1 &\leq& \tilde{M}M(b_0 + c_0) \Delta
\end{eqnarray*} For $4 \tilde{M} M \Delta < 1$, we have,  $a_1 + b_1  + c_1  \leq a_0 + b_0 + c_0 $,
Using $(b_k + c_k) \leq 2\tilde{M}M \Delta (b_{k-1} + c_{k-1}) \leq (2\tilde{M}M \Delta)^k (b_0 + c_0)$.

Similarly,

$| a_k - a_{k-1}|_0 \leq (2\tilde{M}M \Delta)^k (b_0 + c_0)$

Note, $(a_k, b_k, c_k)$ is a Cauchy sequences which converges to $(a_\infty, 0, 0)$, where

$$ | a_{\infty}-a_0 |_0 \leq (b_0 + c_0) \sum_{k=1}^{\infty} (2\tilde{M}M \Delta)^k \leq \frac{2 M \tilde{M}\Delta(b_0 + c_0)}{1 -2\tilde{M}M \Delta} \leq C \Delta, $$ where $C = 4\tilde{M}M (b_0 + c_0)$.

The above excercise was illustrative. Now we use an iterative procedure as above to show Eq. (\ref{eq:iterative.a.increment1}).

Writing $$ Ad_{\bar{K}}(X_d) = \underbrace{P(Ad_{\bar{K}}(X_d))}_{a_0} + \underbrace{Ad_{\bar{K}}(X_d)^{\perp}}_{b_0}, $$
where $a_0 \in \f$ and $b_0 \in \f^{\perp}$, consider again the iterations

\begin{eqnarray*}
U_0 &=& \exp(- \bar{c}_0 \Delta)\exp(a_0 \Delta + b_0 \Delta) \exp(-b_0 \Delta + \bar{c}_0 \Delta) \\
    &=&  \exp(- \bar{c}_0 \Delta) \exp(a_0 \Delta + \bar{c}_0 \Delta + b_0' \Delta^2) \\
    &=& \exp(a_0 \Delta + b_0' \Delta^2 + c_0' \Delta^2)\\
    &=& \exp(a_1 \Delta + b_1 \Delta + c_1 \Delta )
\end{eqnarray*}

We refer to remark \ref{rem:stabilizer}, Eq. \ref{eq:nghdbnd}. Given $b_0 \Delta \in \p$ such that $b_0 \Delta \in \f^{\perp}$. If $A k' A' = -b_0 \Delta + \bar{c}_0 \Delta$,
then $\| k' \| \leq h \| b_0 \Delta \| $ (killing norm).

$\bar{c}_0 \in \k$, is bounded $\bar{c}_0 \leq Mh b_0$, where $M$ as before converts between two different norms.
Using bounds derived above $b_0' \leq \tilde{M}(Mh + 1)b_0$, and $c_0' \leq \tilde{M}Mhb_0$,
$2\tilde{M}(Mh+1) \Delta < 1$, we obtain

which gives $a_0 + b_0' \Delta + \bar{c}_0  \leq a_0 + b_0(\tilde{M}(Mh+ 1)\Delta + Mh) \leq 1 $. For appropriate $M'$, we have

\begin{eqnarray*}
a_1 &\leq& a_0 + \frac{M'}{3} (b_0 + c_0)\Delta \\
b_1 &\leq& \frac{M'}{3}(b_0 + c_0) \Delta \\
c_1 &\leq& \frac{M'}{3}(b_0 + c_0)\Delta
\end{eqnarray*}

we obtain

$$ a_1 + b_1 + c_1  \leq a_0 + M'(b_0 + c_0)\Delta \leq a_0 + b_0 + c_0 $$

where $\Delta$ is chosen small.

\begin{eqnarray*}
U_1 &=& \exp(-(c_1 + \bar{c}_1) \Delta)\exp(a_1 \Delta + b_1 \Delta + c_1 \Delta) \exp(-b_1 \Delta + \bar{c}_1 \Delta) \\
    &=&  \exp(-(c_1 + \bar{c}_1) \Delta) \exp(a_1 \Delta + (c_1 + \bar{c}_1) \Delta + b_1' \Delta^2) \\
    &=& \exp(a_1 \Delta + b_1' \Delta^2 + c_1' \Delta^2) \\
    &=& \exp(a_2 \Delta + b_2 \Delta + c_2 \Delta)
\end{eqnarray*}

Where $\bar{c}_1 \in \k$, such that $\bar{c}_1 \leq Mh b_1$.

Where, using bounds derived above $b_1' \leq \tilde{M}(Mh + 1)b_1$, and $c_1' \leq \tilde{M}(Mhb_1 + c_1)$,
where using the bound $2\tilde{M}(Mh+1) \Delta < 1$, we obtain

which gives $a_1 + b_1' \Delta + (c_1 + \bar{c}_1 ) \leq a_1 + ((1 + Mh)b_1 + c_1) \leq a_0 + b_0 + c_0 $.

We can decompose, $(b_1' + c_1')\Delta^2$, into subspaces $(a_1'' + b_2 + c_2)\Delta$, where
$a_1'' \leq M (b_1' + c_1') \Delta$, $b_2 \leq M(b_1' + c_1')\Delta $ and $c_2 \leq M(b_1' + c_1')\Delta$, where $M$ as before converts between two different norms.

This gives

\begin{eqnarray*}
a_2 &\leq& a_1 + 4\tilde{M}M^2h (b_1 + c_1)\Delta \\
b_2 &\leq& 4\tilde{M}M^2h(b_1 + c_1)\Delta \\
c_2 &\leq& 4\tilde{M}M^2h(b_1 + c_1)\Delta
\end{eqnarray*} For $x = 8 \tilde{M} M^2h \Delta < \frac{2}{3}$, we have,  $a_2 + b_2 + c_2  \leq a_1 + (b_1 + c_1) \leq a_0 + b_0 + c_0$,

Using $(b_k + c_k) \leq x (b_{k-1} + c_{k-1}) \leq x^k (b_0 + c_0)$.

Similarly,

$| a_k - a_{k-1}|_0 \leq x^k (b_0 + c_0)$

Note, $(a_k, b_k, c_k)$ is a Cauchy sequences which converges to $(a_\infty, 0, 0)$, where

$$ | a_{\infty}-a_0 |_0 \leq x (b_0 + c_0)\sum_{k=0}^{\infty} x^k \leq \frac{x (b_0 + c_0)}{1-x} \leq C \Delta, $$ where $C = 16 \tilde{M}M^2 h(b_0 + c_0)$.

From Eq. (\ref{eq:iterative.a.increment1}),

$$ \exp((K_1' Ad_K(X_d) K_1)\Delta) = \exp(-k_1'') \exp(a_0 \Delta + C \Delta^2) \exp(-A k_2'' A'). $$
where $a_0 \Delta + C \Delta^2 \in \f$. By using a stabilizer $H_1, H_2$, we can rotate them to $\a$ such that
$$  \exp(Ad_K(X_d) \Delta)K_1 A K_2 =  K_a H_1 \exp(a_0' \Delta + C' \Delta^2) A H_2 K_b $$ such that $H_1^{-1}(a_0 \Delta + C \Delta^2)H_1 =  a_0' \Delta + C' \Delta^2$ is in $\a$ and  $a_0' = P(H_1^{-1} a_0 H_1)$ is projection onto $\a$ such that

$$ P(H_1^{-1} a_0 H_1) = \sum _k \alpha_k \W_k (X_d).  $$ This follows because the orthogonal part of  $Ad_{\bar K}(X_d)$ to $\f$ written as $Ad_{\bar K}(X_d)^\perp$ remains orthogonal of $\f$

$$ \langle H^{-1} Ad_K(X_d)^{\perp}H, \a \rangle =  \langle Ad_K(X_d)^{\perp}, H \a H^{-1}\rangle = \langle Ad_K(X_d)^{\perp}, \a'' \rangle = 0$$  ($a'' \in \f$),  remains orthogonal to $\a$. Therefore $ P(H_1^{-1} a_0 H_1) = P(H_1^{-1} Ad_{\bar K}(X_d) H_1) =  \sum _k \alpha_k \W_k (X_d)$.

$$ \exp(Ad_K(X_d) \Delta)K_1 A K_2  = K_a \exp(a + a_0' \Delta + C' \Delta^2) K_b. $$

\begin{lemma}\label{lem:join1}{\rm
Given $P = K_1 \underbrace{\exp(a + a_1 \Delta)}_{A_1} K_2 = K_3 \underbrace{\exp(b - b_1 \Delta)}_{A_2} K_4$, where $a, b, a_1, b_1 \in \a$. We can express $$ \exp(b) = K_a \exp(a + a_1 \Delta + {\mathcal W}(b_1) \Delta) K_b, $$ where $\W(b_1)$ is Weyl element of $b_1$. Furthermore

$$ \exp(b + b_2 \Delta) =  K_a'' \exp(a + a_1 \Delta + {\mathcal W}(b_1) \Delta + {\mathcal W}(b_2) \Delta ) K_b''. $$

Note, $A_2 = K_3^{-1} P K_4^{-1}$, commutes with $b_1$. This implies

$A_2 = K \exp(a + a_1 \Delta)K^{-1} \tilde{K} $ commutes with $b_1$. This implies that
$\tilde{K}^{-1} \exp(-Ad_K(a + a_1 \Delta)) b_1 \exp(Ad_K(a + a_1 \Delta)) \tilde{K} = b_1$, which implies that
$\exp(-Ad_K(a + a_1 \Delta)) b_1 \exp(Ad_K(a + a_1 \Delta)) \in \p $. Recall, from remark \ref{rem:stabilizer},

$$ \exp(-Ad_K(a + a_1 \Delta)) b_1 \exp(Ad_K(a + a_1 \Delta)) = \sum_k c_k (Y_k \cos(\lambda_k) + X_k \sin(\lambda_k)), $$

This implies $\sum_k c_k \sin(\lambda_k) X_k = 0$, implying $\lambda_k = n \pi$. Therefore,

$$ \exp(-2 Ad_K(a + a_1 \Delta)) b_1 \exp(2 Ad_K(a + a_1 \Delta)) = b_1. $$

We have shown existence of $H$ such that $H^{-1} b_1 H \in Ad_K(\a)$,

Therefore, $$ \exp(b_1 \Delta) \exp(Ad_K(a + a_1 \Delta)) \tilde{K} = K_a \exp(a + a_1 \Delta + {\mathcal W}(b_1) \Delta) K_b. $$ Applying the theorem again to
$$ \exp(b_2 \Delta)  K_a \exp(a + a_1 \Delta + {\mathcal W}(b_1) \Delta) K_b = K_a'' \exp(a + a_1 \Delta + {\mathcal W}(b_1) \Delta +  {\mathcal W}(b_2) \Delta  ) K_b''. $$
}

\end{lemma}

\begin{lemma}\label{lem:join2}
{ \rm Given  $P_i = K_1^i A^i K_2^i = K_1^i \exp(a^i) K_2^i $, we have $P_{i, i+1} = \exp(H_i^+ \Delta_i^+) P_i$,
and $P_{i, i+1} = \exp(-H_{i+1}^- \Delta_{i+1}^-) P_{i+1}$, where $H_i^+ = Ad_{K_i}(X_d)$. From above we can express

$$ P_{i, i+1} = K_a^{i+} \exp(a^i + a_1^{i+} \Delta_{+}^i + a_2^{i+} (\Delta_{+}^i)^2) K_b^{i+}. $$

where $a_1^{i+}$ and $a_2^{i+}$ are first and second order increments to $a_i$ in the positive direction. The remaining notation is self explanatory.

$$ P_{i, i+1} = K_a^{(i+1)-} \exp(a^{i+1} - a_1^{(i+1)-} \Delta_{-}^{i+1} - a_2^{(i+1)-} (\Delta_{-}^{i+1})^2) K_b^{(i+1)-}. $$

$$ \exp(a^{i+1}) = K_1 \exp(a^i + a_1^{i+} \Delta_{+}^i + a_2^{i+} (\Delta_{+}^i)^2 + {\mathcal W} ( a_1^{(i+1)-} \Delta_{-}^{i+1} + a_2^{(i+1)-} (\Delta_{-}^{i+1})^2 )) K_2. $$

$$ \W(a_1^{(i+1)-} \Delta_{-}^{i+1} + a_2^{(i+1)-} (\Delta_{-}^{i+1})^2 ) = \P (\W(a_1^{(i+1)-})) \Delta_{-}^{i+1} + \P( \W(a_2^{(i+1)-})) (\Delta_{-}^{i+1})^2 = \sum_k \alpha_k \W_k (X_d) \Delta_{-}^{i+1} + o((\Delta_{-}^{i+1})^2 )$$

where, $a^i, a_1^i, a_2^i \in \a$.

}\end{lemma}

Using lemma \ref{lem:join1} and \ref{lem:join2} , we can express

$$P_n(T) = K_1 \exp(a_n) \exp K_2 = K_1 \exp(\sum_i \W(a_i^{+}) \Delta_i^{+} + \W(a_{i+1}^{-})\Delta_{i+1}^{-})\exp(\underbrace{\sum o(\Delta^2)}_{\leq \epsilon T}) K_2 $$

Letting $\epsilon$ go to $0$, we have

$$ P_n(T) = K_1  \exp(T \sum_i \alpha_i \W_i(X_d)) K_2. $$

Hence the proof of theorem \ref{th:generalmodel}. {\bf q.e.d.}

\begin{corollary}\label{cor:generalmodel}{\rm Given $U$, in compact Lie group $G$, with $X_d, X_j$ in its Lie algebra $\g$. Given the Cartan decomposition $\g = \p \oplus \k$, where $X_d \in \a \subset \p$
\begin{equation}
\dot{U} = ( X_d + \sum_j u_j(t) X_j ) U,  \ \ U(0) = \Iden,
\end{equation} and $\{X_j\}_{LA} = \k$. The elements of the reachable set at time $T$, takes the form
$ U(T) \in$  $$ S = K_1 \exp (T \sum_k \alpha_k \ \W_k X_d \W_k^{-1}) K_2, $$ where $\W_k$ are Weyl elements and $K_1, K_2, \W_k \in \exp(\k)$. $S$ belongs to the closure of reachable set. }\end{corollary}

\begin{theorem}{\bf Co-ordinate theorem}\label{th:coordinate}{\rm \ \ Let $\g = \p \oplus \k$ be a Cartan decomposition with Cartan subalgebra $\a \in \p$. Let $a \in \a$ be a regular element such that $\f = \a$. Given ${\rm ad}_a^2: \p \rightarrow \p$ symmetric. Let $Y_i$ be the eigenvectors of ${\rm ad_a}^2$ that are orthogonal to $\a = \{ Z_j \}$. Let $X_i = \frac{[a, Y_i]}{\lambda_i}$, $\lambda_i > 0$, where $-\lambda_i^2$ is a eigenvalue of $ad_a^2$. Then $\k = \{X_i \} + \k_0 =  \{X_k \}$, where $[a,  \k_0]=0$ and $X_i \perp \k_0$.
$$ ad_a(Y_i) = \lambda_i X_i, \ \ ad_a(X_i) = -\lambda_i Y_i $$
\begin{equation}
A X_i A^{-1} = \cos(\lambda_i) X_i - \sin(\lambda_i) Y_i,
\end{equation}where $A = \exp(a)$.

Given $U = K_1 \exp(a) K_2$, consider the map

$$ U(a_i, b_j, c_k) =  \exp(\sum_k c_k X_k) \ K_1  \exp(\sum_j b_j Z_j) \ \exp(a) \ \exp(\sum_i a_i X_i)K_2 $$
such that $U(0, 0, 0) = U$.
\begin{eqnarray}
\frac{\partial U}{\partial a_i}|_{(0,0,0)} &=& (\cos(\lambda_i) Ad_{K_1}(X_i) - \sin(\lambda_i) Ad_{K_1}(Y_i))\ U \\
\frac{\partial U}{\partial b_j}|_{(0,0,0)} &=& Ad_{K_1}(Z_j)\ U \\
\frac{\partial U}{\partial c_k}|_{(0,0,0)} &=& X_k \ U.
\end{eqnarray}

$Y_i, Z_j$ span $\p$, $Ad_{K_1}(Y_i)$, $Ad_{K_1}(Z_j)$, span $\p$. $Ad_{K_1}(Y_i)$, $Ad_{K_1}(Z_j)$, $X_k$ span $\p \ \oplus \ \k$. $\cos(\lambda_i) Ad_{K_1}(X_i) - \sin(\lambda_i) Ad_{K_1}(Y_i)$, $Ad_{K_1}(Z_j)$ and $X_k$, span $\p \oplus \k$.

By inverse function theorem $U(a_i, b_j, c_k)$ is a nghd of $U$, any curve
$U(t)$ passing through $U$, at $t=0$, for $t \in (-\delta, \delta)$ can be written as

$$ U(t) =  \exp(\sum_k c_k(t) X_k) K_1  \exp(\sum_j b_j(t) Z_j) \exp(a) \exp(\sum_i a_i(t) X_i) K_2 = K_1(t)A(t)K_2(t). $$
$(a_i, b_j, c_k)$ are coordinates of nghd of $U$.

Given $U(0) = K_1 \exp(a) K_2$ such that $a$ is regular ($\f = \a$, see remark \ref{rem:stabilizer}), we can represent a curve $\dot{U}(t) = Ad_{K(t)}(X_d) U(t)$ 
passing through $U(0)$ as $U(t) = K_1(t)A(t)K_2(t)$, where
$\dot{K_1} = \Omega_1(t) K_1$, $\dot{K_2} = \Omega_2(t) K_2$ and $\dot A(t) = \Omega(t) A(t)$ where $\Omega_1(t), \Omega_2(t) \in \k$ and  
$\Omega(t) \in \a$. Differentiating, we get 

$$ Ad_{K(t)}(X_d)  U(t) = (\Omega_1 + K_1 \Omega K_1^{-1} + K_1 A \Omega_2 A^{-1} K_1^{-1})U(t), $$
which gives for $\bar{K} = K_1^{-1} K$, and $\Omega_1' = K_1^{-1} \Omega_1 K_1 \in \k$, 

$$ Ad_{\bar K}(X_d) = \Omega_1' + \Omega + A \Omega_2 A^{-1}. $$ Using $A \Omega_2 A^{-1} \perp \a$, we obtain $\Omega = P(Ad_{\bar K}(X_d))$, projection
of $Ad_{\bar{K}}(X_d)$ on $\a$. $A(t)$ evolves as this projection, which lies in convex hull of Weyl points of $X_d$ by Kostant Convexity theorem.}
\end{theorem}

\section{Roots and reflections}

\begin{remark}{\rm {\bf Roots:} Let $\g$ be real, compact, semisimple Lie algebra, with negative definite killing form $\langle ., . \rangle$. Let $E_i$ be basis of $\g$, orthonormal, wrt to the killing form. ${\rm ad}_X$ is skew symmetric matrix, wrt to these basis.

$$ \langle E_i, {\rm ad}_X(E_j) \rangle  = tr({\rm ad}_{E_i}{\rm ad}_{[X, E_j]}) = tr({\rm ad}_{E_i} [{\rm ad}_X {\rm ad}_{E_j}]) = -  \langle E_j, {\rm ad}_X(E_i) \rangle.  $$
where, we use, $ad_{[X, Y]} = [ad_X, ad_Y]$, which follows from Jacobi identity, $[[x, y], z] = [x [y, z]] - [y [x, z]]$, $ad_{[X, Y]} = [ad_X, ad_Y]$.}
\end{remark}

Let $a \in \a$. Eigenvalues of $A = ad_{a}$, are imaginary ($A$ is skew symmetric), as $Ax = \lambda x$, implies $\lambda = \frac{x'Ax}{x'x}$, implying,
$\lambda^{\ast} = -\lambda$. The coefficients of characteristic polynomial being real, the roots, occur in conjugate
pair. $(A - \lambda I)^2 x = 0$, implies, $-((A - \lambda I)x)'(A-\lambda I)x = 0$, implying
$(A- \lambda I)x = 0$. Repeated use of this gives,
$(A - \lambda I)^{k} x =0$, implies $(A - \lambda I) x = 0$, hence $A$ diagonalizable. If $\lambda_i \neq \lambda_j$, $A x_i = \lambda_i x_i$, implying $x_j' A x_i = \lambda_i x_j' x_i$, implying $\lambda_j x_j'x_i = \lambda_i x_j' x_i$, implying $x_j'x_i = 0$.

Given, $A(x + i y) = i \lambda (x + i y)$, let,
$x = x_p + x_k$ and $y = y_p + y_k$, be direct decomposition in $p + k$ parts.
\begin{equation}
A ( x_p + x_k + i (y_p + y_k)) = i \lambda (x_p + x_k + i(y_p + y_k)); \\
\end{equation}
\begin{equation}
A x_p = -\lambda y_k ;  \ \ A y_k = \lambda x_p;
\end{equation}
\begin{equation}
Ax_k = - \lambda y_p ; \ \ Ay_p = \lambda x_k.
\end{equation}

\begin{equation}
A ( x_p + i y_k ) = i \lambda (x_p + i y_k); \ \ A (y_p - ix_k) = i \lambda (y_p - ix_k)
\end{equation}

Eigenvectors of $A$, have the form $x_p \pm i y_k$, with conjugate eigenvalues. Choose a basis for $\a$ as $a_i$, with $A_i = {\rm ad}_{a_i}$. Since $A_i$, commute, we have $A_1 A_2 x = A_2 A_1 x = \lambda A_2 x$, where, $\lambda$, is a an eigenvalue of $A_1$. If $\lambda$ is a distinct eigenvalue, $A_2 x = \mu x$, $x$, is a eigenvector of $A_2$. If $\lambda_k$ has multiplicity $m$ with eigenvectors $x^k_1, \dots, x^k_m$, with $A_2 x^k_i = \sum_{j=1}^m C_{ij}x^k_j$ . Let $X^k = [x^k_1, x^k_2, \dots, x^k_{m}]$, where eigenvectors have been stacked as columns, $A_2 X^k  = X^k C$. Let $\alpha$, be an eigenvalue of $C$, then
$(C - \alpha I)y = 0$, this means, $(A_2 - \alpha I)X^k y = 0$, which implies $\alpha$ is an eigenvalue of $A_2$,
and hence imaginary. Furthermore, $(C - \alpha I)^d y = 0$ implies $(A_2 - \alpha I)^d X^k y = 0$. This entails,
$(A_2 - \alpha I)X^k y = 0$ and hence $(C - \alpha I)y = 0$. Therefore, $C = U \Sigma U^{-1}$ can be diagonalized.
Let $Y^{kl} = [y^{kl}_1, \dots, y^{kl}_d]$, is a subset of columns of $X^k U$, with eigenvalues, $\lambda_k, \lambda_l$, for $A_1, A_2$ respectively. This process can be continued. Let ${\cal I} = (\lambda_{i_1}, \dots, \lambda_{i_n})$ be a multi-index, such that $Y^{\cal I}$, be the set of eigenvectors with eigenvalues $(\lambda_{i_1}, \dots, \lambda_{i_n})$ for $A_1, \dots, A_n$ respectively. Then $Y^{\cal I}$ is $\perp$ to $Y^{\cal I'}$ where ${\cal I} \neq {\cal I'}$.
Finally a column of $Y^{\cal I}$, can be written as $(x_p + x_k + i (y_p + y_k))$. Let $(x_p + i y_k)_s$, be
independent vectors distilled from columns of $Y^{\cal I}$, denoted as $\tilde{Y}^{\cal I}$. (Note, if $(x_p + i y_k)$, are independent under reals, they are independent under complex). Then
$\tilde{Y}^{\cal I}$, has same number of columns as $Y^{\cal I}$, as $\tilde{Y}^{\cal I}$, and $Y^{\cal I}$ are independent, and $Y^{\cal I} = \tilde{Y}^{\cal I} U$ and by
Jordan normal form, $(x_p + i y_k)_s$ can be expressed as linear combination of columns of $Y^{\cal I}$.
The $x_p$ corresponding to distinct ${\cal I}$ (modulo $-{\cal I}$) are orthogonal, as $x_p$, is a eigenvector of
$A^2 x_p = -\lambda^2 X_p$, and since $A^2$, is symmetric, $x_p$ corresponding to distinct $\lambda$ are perpendicular.

If $x + i y$, is a zero eigenvector of $A_i$, then $A_ix = A_iy = 0$. Let $\I_0$ correspond to multi-index, with
eigenvalues identically zero. Then $ Y^{{\cal I}_0} = \{x_{1}, \dots, x_{j}, y_{1}, \dots, y_{k} \}$,
where $x_{j} \in \p$ and $y_k \in \k$. Given $x \in \p$, $x \in Y^{{\cal I}_0}$, iff $x \in \a$, as $\a$,
the maximal abelian subspace $\in \p$.

These eigenvectors ${\cal I}$ can be stacked as a Matrix $J$, which simultaneously diagonalizes all $A_i$, i.e.,
$JA_iJ^{-1} = \Sigma_i$. For ${\cal I} \neq {\pm \cal I'}$, if $(x_p + i y_k) \in {\cal I}$ and $(x_p' + iy_k') \in
{\cal I'}$, then $x_p \perp x_p'$ and $y_k \perp y_k'$. If $(x_p + iy_k) \in {\cal I}$, then $(x_p - iy_k) \in {- \cal I}$. We abbreviate $x_p + i y_k$ as $p + i k$ and call them {\it \bf roots}. We also use the notation $k + ip$ for roots, obtained by multiplication by $i$.

We use roots to show existence of a {\it regular element}.
Given roots $p_j + i k_j$. $p_j$ span ${\a^{\perp}}$.
Let $X_i$ be a basis for $\a$. Then $X_i, p_j$ forms a complete basis for $\p$.
Consider the matrix $Z$, such that $Z_{ij} k_j = [X_i, p_j]$. We form the ratio,
$\alpha_{n-1} = \min_j \{ \frac{|Z_{nj}|}{|Z_{(n-1)j}|} \}$, where both numerator and denominator, are non-zero. When, no such pair exits $\alpha_{n-1} = 1$. We $X_{n-1}$ to $X_{n-1} \leftarrow X_{n} + \frac{\alpha_{n-1}}{2} X_{n-1}$.
Similarly define $\alpha_{k}$. Let $X_{k} \leftarrow X_{k+1} +
\frac{\alpha_{k}}{2} X_{k}$. Then $X_{1}$ as formed from linear
combination $X_i$ is a {\it regular element}.

Given a root vector $e = p + ik$, (with $p, k$ normalized to killing
norm $1$) its value $\alpha$ defined as $[a,
  p + ik ] = -i \alpha(a) (p + ik)$, can be read by taking inner
product with vector $[p, k] \in \a$.
$$ \langle a, [p, k] \rangle = \langle [a, p], k \rangle = \alpha. $$
We represent the root by its representative vector $e = [p, k] \in \a$.
Choose a basis for the roots $e_k$. We can express all roots in terms
of $e_k$ as coefficients $(c_1, \dots, c_k, \dots, c_n)$. The ones
with positive leading non-zero entry are called positive and
viceversa.

\begin{theorem}\label{rem:reflection}{\rm {\bf Reflection:} $\g = \p \oplus \k $ , Let $Y_\alpha \pm i X_\alpha$, where that $Y_\alpha \in \p$ and $X_\alpha \in \k$, are the roots, such that,

$$ [\a, Y_{\alpha}] = \alpha(\a) X_{\alpha} ; \ \ [\a, X_{\alpha}] = - \alpha(\a) Y_{\alpha} $$

$$ [\a, Y_{\alpha} + i X_{\alpha} ] = -i \alpha(\a) ( Y_{\alpha} + i X_{\alpha}) $$

Note, $[X_{\alpha}, Y_{\alpha}] \in \a$. Note, $[X_{\alpha}, Y_{\alpha}] \in \p$, let $a_0 \in \a$ be regular.

$$ ad_{a_0}([X_{\alpha}, Y_{\alpha}]) = [ad_{a_0}(X_{\alpha}), Y_{\alpha}] + [X_{\alpha}, ad_{a_0}(Y_{\alpha})] = 0. $$

Observe,

$$ \exp(s X_{\alpha})[X_{\alpha}, Y_{\alpha}] \exp(-s X_{\alpha}) = [X_{\alpha}, Y_{\alpha}] + \beta s Y_{\alpha}
+ \frac{\beta s^2}{2}[X_{\alpha}, Y_{\alpha}] + \dots $$

where $\beta < 0$. The above expression can be written as,

$$ \exp(s X_{\alpha})[X_{\alpha}, Y_{\alpha}] \exp(-s X_{\alpha}) = \cos (\sqrt{|\beta|} s) [X_{\alpha}, Y_{\alpha}] - \sqrt{|\beta|} \sin(\sqrt{|\beta|}s) Y_{\alpha}. $$

By choosing, $s = \frac{\pi}{\sqrt{|\beta|}}$, we have

$$ U [X_{\alpha}, Y_{\alpha}] U^{-1} = \exp(s X_{\alpha})[X_{\alpha}, Y_{\alpha}] \exp(-s X_{\alpha}) = -[X_{\alpha}, Y_{\alpha}]. $$

Given

$$ Z = c [X_{\alpha}, Y_{\alpha}] + \sum_k \alpha_k Z_k, $$ where $\langle [X_{\alpha}, Y_{\alpha}], Z_k \rangle = 0$.

This implies that $[Z_k, X_{\alpha}] = 0 $, else $[Z_k, X_{\alpha}] = \alpha(Z_k) Y_{\alpha}$. Since

$$ \langle Y_{\alpha}, [Z_k, X_{\alpha}] \rangle = \langle Z_k, [X_{\alpha}, Y_{\alpha}] \rangle = 0, $$

implying $\alpha(Z_k) = 0$. This implies for $U  = \exp( \frac{\pi}{\sqrt{|\beta|}} X_{\alpha})$

$$ U Z U^{-1} = - c [X_{\alpha}, Y_{\alpha}] +  \sum_k \alpha_k Z_k. $$

This is reflection in the plane given by $\alpha(.) = \langle [X_{\alpha}, Y_{\alpha}], . \rangle = 0$.

In orthonormal basis $E_i$ (for $\a$),  $Z$, $[X_{\alpha}, Y_{\alpha}]$ and $Z_k$, takes the form of coordinates, $\z, \m, \z_k$, respectively,
where, $\z = c \m + \sum_k \z_k$, and $\z_k \perp \m$, the reflection formula takes the form

$$ R_{\m} (\z) = \z - 2 \frac{\langle \m, \z \rangle}{\langle \m, \m \rangle}\m  = -c \m + \sum_k \z_k. $$ }
\end{theorem}

\begin{remark}
{\rm When $\a$ is one-dimensional in theorem \ref{th:generalmodel}, we can choose $U$ as in above remark \ref{rem:reflection}, such that $U X_d U^{-1} = -X_d$. Let $X(T) \in KU_F$, belong to coset of $U_F$, where $\dot{X} = Ad_K(X_d)X$. Let the length $L(X(t)) = \beta T$, where $\beta = |Ad_K(X_d)|$. Form of geodesics say that we have for $l \leq \beta T$ such that $\exp(Yl) \in KU_F$ , where $|Y| = 1$. Therefore, $\exp((\beta Y) \frac{l}{\beta}) \in KU_F$. Let $Ad_K(X_d) = \beta Y$, by
appropriate choice of $K$. This is achieved by Maximization of $\langle Ad_K(X_d), Y \rangle$, w.r.t $K$, which yields $[Ad_K(X_d), Y] =0$. This, gives
$Ad_K(X_d) = \pm \beta Y$. We can choose either, by the choice of $U$. Therefore, $X(T) = K_1 \exp(t X_d) K_2$, where
$t = \frac{l}{\beta} \leq T$. For $t < T$, we can use $U$, to insure $U_F = K_1 \exp(T (\alpha X_d + (1-\alpha) U X_d U^{-1}) )K_2$. We get the form of the reachable set in  theorem \ref{th:generalmodel}, by a geodesic argument.}
\end{remark}

\begin{remark}\label{rm:Weyltransitive}{\rm Let $e_j^{+} = k_j + i p_j$, be positive
    roots. The roots divides $\a$, into connected regions called Weyl chambers defined by $sign(e_j^{+}(x)) = \pm 1$, where the signs donot change over a connected region. On the boundary of a Weyl chamber, some of $e_j^{+}(x) = 0$.  By a sequence of reflections $s_j$, around roots
$e_j^{+}$, we can map one Weyl chamber into another. Let $x$ be a point in Weyl chamber $A$. Choose a point $y$ in the interior of principal Weyl chamber $\c$ defined as $e_j^{+}>0$,
choose a $e_k^{+}$ such that $e_k^{+}(x) < 0$. We can decompose $y =
y^{\parallel} + y^{\perp}$, similarly for $x = x^{\parallel} +
x^{\perp}$, where $\perp$ and $\parallel$ is w.r.t. the hyperplane of 
$e_k^+$. Then the distance between
$x$ and $y$ is $d_1 = \sqrt{|y^{\parallel} - x^{\parallel}|^2 + (|x^{\perp}| + |y^{\perp}|)^2}$,
after reflection the distance changes to
$d_2 = \sqrt{| y^{\parallel} - x^{\parallel}|^2 + (|x^{\perp}| - |y^{\perp}|)^2}$, as part
parallel to hyperplane of $e_k^{+}$ is invariant under reflection.

$$ d_1^2 - d_2^2 = 4 |y^{\perp}||x^{\perp}| = 4 | \langle y,\ e_k^{+}
\rangle \langle x,\ e_k^{+} \rangle|. $$ Let $y_{min}$ be the minimum of
$|\langle y,\ e_j^{+} \rangle|$, and $x_{min}$ be the minimum of
nonzero $|\langle x,\ e_j^{+} \rangle|$ taken over all $j$.

Let $\Delta = 4 x_{min} y_{min}$.


We can continue this process by finding next $k$, such that, where $ sign(e_k^{+}(x))\neq sign(e_k^{+}(y))$.

The value of root on reflected $x$, can be evaluated by permuting roots and evaluating them on original $x$. Let $\W$, be the Weyl rotation corresponding to reflection
$s$, then

$$[\W \a \W^{-},\ k + ip] = \lambda(\W \a \W^{-}) (k + ip) $$

$$ \W  [ \a, \W^{-} (k+ ip) \W ] \W^{-1} = \lambda(\W \a \W^{-}) (k + ip) $$

$$ [ \a, \W^{-} (k+ ip) \W ]  = \lambda(\W \a \W^{-}) \W^{-} (k + ip) \W $$

Thus $\W^{-} (k+ ip) \W = k_1 + i p_1$ is a root, such that its value $\lambda_1(\a)$
at $\a$ is same as $\lambda(\W \a \W^{-})$ abbreviated as  $\lambda(\W \a)$.

$$[\W_n \dots \W_1 \a \W_1^{-} \dots \W_n^{-}, k + ip ] = \lambda(\W_n \dots \W_1 \a ) (k + ip). $$

$$[\a, \W_1^{-} \dots \W_n^{-} k + ip \W_n \dots \W_1 ] = \lambda(\a ) \W_1^{-} \dots \W_n^{-} k + ip \W_n \dots \W_1. $$

$ \langle x e_k^{+} \rangle \geq x_{min} $, where $x_{min} = min_j \{ | \langle x e_j^{+} \rangle| \neq 0 \} $. Each reflection reduces the squared distance by atleast
$\Delta$. In finite rotations, either the distance is reduced to $0$ or $x$ reaches principal Weyl chamber, $\c$. If $x$ is an interior point then after reflections it stays an interior point. Boundary points
go to boundary points. For $y \in \c$, if $\W(y) \in \c$, then $\W$ is identity. Weyl rotations act simple on Weyl chambers \cite{Helg}. Simple action entails that any $\W$ can be written as product of finite reflections.} \end{remark}

\begin{remark}{\rm Given the reflection formula of root $\alpha$ round $\beta$, 
$$ \alpha \rightarrow \alpha - 2 \frac{\langle \alpha, \beta \rangle }{\langle \beta, \beta \rangle} \beta. $$ We claim $2 \frac{\langle \alpha, \beta \rangle }{\langle \beta, \beta \rangle}$ is an integer.
Consider the root $\beta = p + i k$, and $e = \frac{p + ik}{\|[p, k]\|}$ and $f = \frac{p - ik}{\|[p, k]\|}$ and $h = [f, e]$ where $h = \frac{2 i [p, k]}{\|[p, k]\|^2}$. Then
$[h, e] = 2 e$ and $[h, f] = -2 f$. Consider the root $\upsilon$. Then using the convention $h(\upsilon) = [h, \upsilon]= \lambda_0 \upsilon $, we have
$h(e (\upsilon)) = e (h (\upsilon)) + [h, e] (\upsilon) = (\lambda_0 + 2) e(\upsilon)$. In general then, $h(e^k (\upsilon)) = (\lambda_0 + 2k) e^k (\upsilon)$. $e$ is called the {\it raising operator}. Now let $e^{k+1}(\upsilon)  = e (w) = 0$, where $h (w) = \lambda w =  (\lambda_0 + 2k) w$. Now consider $w, f (w), \dots f^d (w) $ such that $f^{d+1} (w) = 0$.
$h(f (w)) = f (h (w)) + [h, f] (w) = (\lambda - 2) f(w)$. In general then, $h(f^k (w)) = (\lambda - 2k) f^k (w)$. $f$ is called the {\it lowering operator}. Furthermore $e(f^k (w))$ lies in span of $w, f(w), \dots, f^{k-1}(w)$. Its true by induction. When $k=1$, we have 
$e (f (w)) = -h w = -\lambda w$. Assuming true for $k$, we have $e(f^{k+1}(w)) = f (e (f^{k}w)) - [f, e](f^{k} w)$. Hence $h = [f, e]$ which is a diagonal matrix on
 $w, f(w), \dots, f^{d}(w)$ is a commutator $[f, e]$, hence must have trace zero. The trace of $h = (\lambda -d)(d + 1)$, hence $\lambda = d$ an integer.
This says that $h(\alpha)$ is a integer for root $\alpha$. If the root $\alpha = p_1 + i k_1$. Then this says that 
$$ h (\alpha) = 2 \frac{\langle [p, k][p_1 k_1] \rangle }{\| [p, k] \|^2} = 2 \frac{\langle \alpha, \beta \rangle }{\langle \beta, \beta \rangle}, $$ is an integer.
Now this says that 
$$ 2 \frac{\langle \alpha, \beta \rangle }{\langle \beta, \beta \rangle} 2 \frac{\langle \beta, \alpha \rangle }{\langle \alpha, \alpha \rangle} = 4 \cos^2 \theta $$
where $\theta$ is the angle between the two roots. This says that $4 \cos^2 \theta$ only takes integer values $\{0, 1, 2, 3 \}$. Hence the angle between the roots can only take values $\{ 0, \frac{\pi}{6}, \frac{\pi}{4},  \frac{\pi}{3},  \frac{5 \pi}{6}, \frac{3 \pi}{4},  \frac{2 \pi}{3}, \pi.\}$}\end{remark}

\begin{remark}{\rm 
There exist a basis $e_i$ for the $\a$ such that all positive roots can be expressed as
$f = \sum \alpha_j e_j$ where $\alpha_j > 0$ are integers. There exits a $z$ such that $\langle z, f \rangle > 0$ for all positive $f$.
Lets collect from $f$, all roots such that cannot be written as a sum of other two roots $\alpha + \beta$, we call this set $\B$, then set of simple positive roots.
We choose from $\B^c$, $x$ such that $\langle z, x \rangle$ is smallest in $\B^c$. Then it follows $x = x_1 + x_2$ and both $x_1$ and $x_2$ are in $\B$.
We claim elements of $\B$ make an obtuse angle among themselves. Suppose not. Then given $\alpha, \beta \in \B$, making an acute angle, reflect around root of larger magnitude, say $\beta$. This gives
$$ \alpha \rightarrow \alpha - 2 \frac{\langle \alpha, \beta \rangle }{\langle \beta, \beta \rangle} \beta. $$ Since $2 \frac{\langle \alpha, \beta \rangle }{\langle \beta, \beta \rangle}$ is a integer, it possible value is $1$. Hence $\alpha -\beta$ is a root. Then $\alpha$ is not a simple root, which is a contradiction. Given
$e_i \in \B$, we claim $e_i$ form an independent set. Suppose dependent then $\sum_i \alpha_i e_i = 0$ for nonzero $e_i$. We can write this as 
$x = \sum_i \alpha_i e_i = \sum_j \beta_j f_j$ where $\alpha_i >0 , \beta_j > 0$. Then $\langle x, x \rangle = \langle \sum \alpha_i e_i,  \sum \beta_j f_j \rangle  \leq 0$.
This implies that $x = \sum_i \alpha_i e_i = 0$. Since $\langle e_i , z \rangle >0$, implies $\alpha_i = 0$ and $\beta_j = 0$. Hence $\B$ is an independent set. Hence the proof that $\B$ forms a basis. From previous remark, possible angles between elements of $\B$ is $90^{\circ}, 120^{\circ}, 135^{\circ}, 150^{\circ}$. We call $\B$ fundamental roots. } 
\end{remark}

\begin{theorem}{\rm Let $\g = \p \oplus \k$ be simple algebra (no ideals). Given any $a_1 \in \a$, We show $\a$ is spanned by ${\W}_i(a_1) = Ad_{k_i}(a_1)$.}\end{theorem}

Consider the reflection around the root $e_1$, where $e_1$ is independent of $a_1$, this gives,
$a_1 \rightarrow a_1 - \langle a_1 e_1 \rangle e_1 = a_2$. $a_2$ is independent of $a_1$. Let $e_{k}$ be independent of the generated vectors $a_1, \dots, a_k$, and
not perpendicular to these, then reflecting these around $e_{k}$, produces $a_{k+1}$, which is independent of these. If no such $e_k$ can be found beyond $k-1$, chain terminates. Then we can divide the root vectors into two categories, $R_1 = \{e_1, \dots, e_q \} \in span \{ a_1, \dots, a_k \} = {\a}_1$ and $R_2 = \{e_{q+1}, \dots, e_N \} \perp span \{ a_1, \dots, a_k \} = {\a}_2$. Given $e \in R_1$ and $f \in R_2$, then
$e \perp f$. $[e, f ]$, if not zero, is $e + f$, has non-vanishing inner product with $e$ and $f$. Then $[e, f]$ is root that is neither parallel or perpendicular to $span\{a_k \}$, therefore $[e, f] =0$

This divides nontrivial roots into two commuting sets $R_1$ and $R_2$. Let $\k_1$,
and $\k_2$ be the $\k$ part of the roots $k + ip$ comprising $R_1$ and $R_2$. Similarly $\p_1$ and $\p_2$.

$$ [\a_1, \k_1] \in \p_1 $$
$$[ \a_1, \p_1 ] \in \k_1 $$
$$[ \a_1, \k_2 ] = 0 $$
$$[ \a_1, \p_2 ] = 0 $$

Similarly, for $\a_2$.

$$ [\k_1, \p_1 ] \in \p_1 \oplus \a_1. $$
$$ [\k_2, \p_2 ] \in \p_2 \oplus \a_2. $$

Follows from $\a_2$ and $\p_2$ commute with $\k_1$ and $\p_1$ and viceversa.
Then
for $k_1 \in {\k}_1$ and $k_2 \in {\k}_2$, $[k_1, k_2]= 0$. This follows from $[k_1 + ip_1, k_2 \pm i p_2] = 0$. Similarly $[{\p}_1, {\p}_2] =0$ and

$$ [\p_1, \k_2 ] = 0 $$
$$ [\k_1, \p_2 ] =  0 $$
$$ [\k_1, \k_1 ] \perp \k_2 $$
$$ [\k_2, \k_2 ] \perp \k_1 $$
$$ [\p_1, \p_1 ] \perp \k_2 $$
$$ [\p_2, \p_2 ] \perp \k_1 $$

 Let $\k_0$ be
trivial roots in $k$, i.e., $[a, \k_0] = 0$.

$$[\k_0, \k_1 ] \in \k_1 $$
$$[\k_0, \p_1 ] \in \p_1 $$

Similarly for $\k_2$, $\p_2$.

Let $B_1 \in \k_0$ be generated by $\k_1$ orthogonal part of $[\k_1, \k_1]$ and $[\p_1, \p_1]$. Let $B_2 \in \k_0$ be generated by $\k_2$ orthogonal part of $[\k_2, \k_2]$ and $[\p_2, \p_2]$. $[B_1, \a_1] =0$, $[B_1, \k_1] \in \k_1$ and $[B_1, \p_1] \in \p_1$,
$[B_1, B_1] \in \tilde{\k}_1$, where, ${\tilde{\k}}_1 = \k_1 \oplus B_1$ and ${\tilde{\k}}_2 = \k_2 \oplus B_2$. Let $B_3$ be part of $\k_0$, orthogonal to $B_1$ and $B_2$.  Note,
$[{\tilde{\k}}_i, {\tilde{\k}}_i] \in {\tilde{\k}}_i $.
${\tilde{\k}}_1 \perp {\tilde{\k}}_2$. $[B_3, \tilde{\k}_i] \in \tilde{\k}_i$.

Then ${\mathfrak I}_1 = \a_1 \oplus \p_1 \oplus {\tilde{\k}}_1$ and ${\mathfrak I}_2 = \a_2 \oplus \p_2 \oplus {\tilde{\k}}_2$, are non-trivial ideals.

Therefore $k =n$, i.e, $span\{a_1, \dots, a_n \} = \a$. We show positive span of
$\{a_1, \dots, a_n \} = \a$. Consider the convex hull of
$C = {\W}_i a_1 $, where ${\W}_i = Ad_{k_i}$. Suppose origin is not in the convex hull. By Hahn Banach theorem we can find a separating Hyperplane such that
$\langle c, x \rangle = \sum c_i x_i > 0$ for all Weyl points of $a_1$. We can write the hyperplane in terms of $n$ independent root vectors as $\sum_j b_j \langle s_j, x \rangle$,  $c = \sum b_j s_j$. Let $y$ be chosen such that $sign \langle s_j, y \rangle = -sign(b_j)$. By reflecting around plane $s_j$, if $sign(\langle s_j, x \rangle ) \neq sign(\langle s_j, y \rangle)$,
we decrease the distance between $a_1$ and $y$ (this is same idea as in remark \ref{rm:Weyltransitive}). In finite steps
$b_j \langle s_j, x \rangle \leq 0$. Therefore $\sum_j b_j \langle s_j, x \rangle \leq 0$. Therefore $0 \in C$, i.e., $\sum_j \alpha_j Ad_{k_j} a_1 = 0$, i.e.,
$-a_1 = \sum \alpha_j Ad_{k_j}(a_1)$. Hence the proof.

\begin{theorem}\label{th:semisimple}{\rm Let $V_i$ be mutually commuting root vectors, such that no further subdivision in commuting sets is possible. Given roots
$e_l = k_l + i p_l \in V_l$ and $e_m = k_m + i p_m \in V_m$,
we have $[k_l \pm i p_l, k_m \pm i p_m ] = 0$ which implies
$[k_l, k_m] = [k_l, p_m] = [k_m, p_l] = [p_l, p_m] = 0$
The associated
root vectors are $[k_l, p_l]$ and $[k_m, p_m]$. The
$$ \langle [k_l, p_l] [k_m, p_m] \rangle = \langle k_l [p_l, [k_m, p_m]] \rangle = 0 $$
where we use Jacobi identity.

Let $\a_i$ be the subspace spanned by root vectors $V_i$,
a direct decomposition of $\a$, into root spaces,
$$ \a = \a_1 \oplus \a_2 \dots \oplus \a_s. $$
Given an element $a \in \a$, we can decompose,
$a = a_1 + a_2 + \dots + a_s $
Reflecting in root vectors in $V_i$, only reflects root vector $a_i$.
Reflecting $a_1$ we can produce $\sum_i \alpha_i {\W}_1(a_1) = -a_1$
leaving $a_i$, $i \neq 1$ invariant. We can synthesize a convex combination that
synthesizes $\pm a_{1}$. Using the construction detailed before, we can synthesize
$V_1$. Similarly we can synthesize all $V_j$, and hence any $V$.

Let $\k_i$ and $\p_i$ be the subspace formed from the $k$ and $p$ parts of the roots
in $V_i$. Then $[\k_i, \p_j] = 0$, $[\k_i, \a_j] = 0$,
$[\k_i, \p_j] = 0$, where $i \neq j$.
We have $[\k_i, \p_i] \perp \p_j$ , $[\k_i, \p_i] \perp \a_j$, for $i \neq j$. This implies
$[\k_i, \p_i] \in \p_i \oplus \a_i $. We have
$[\k_0, a] = 0$ and $[\k_0, \p_i] \in \p_i$. This implies
$\tilde{\p} = \sum_{i = 1}^m \a_i \oplus {\p}_i$, where $m < s$, is invariant under $ad_k$ and $Ad_k$.

Given $X_d \in \tilde{\p}$, the solution to the differential equation

$$ \dot{X} = (X_d + \sum_i u_i k_i) X, \ \ k_i \in \k $$ are confined to the invariant,
manifold $\tilde{G} = \exp(\{\tilde{\p}, \k \})$. $K = \exp(\k)$ is closed subgroup,
we can decompose $\tilde{G} = \exp(\tilde{\p}) K$. Given $Y \in \tilde{\p}$, we can rotate it to Cartan subalgebra $\a$, i.e.,
$Ad_K(Y) \in \a$. Since $Y \in \tilde{\p}$ is $Ad_K$ invariant,
$Ad_K(Y) \in \sum_{i=1}^m \a_i$.
$\tilde{G} = K_1 \exp(\sum_{i=1}^m b_i)K_2$, where $b_i \in \a_i$.
We can synthesize $\sum_{i}^m b_i = \sum_j \alpha_j Ad_{k_j} X_d$
as detailed before.
$\tilde{G} = K_1 \exp(\sum_j \alpha_j Ad_{k_j} X_d)K_2$.}
\end{theorem}


\begin{theorem}{\rm Given $\g = \p \oplus \k$, Let $a \in \a$. The number of Weyl points $\W a \W \in \a$ are finite.}\end{theorem}

{\bf Proof:} For $j = 1, \dots, m$, we choose as basis of $\g$,
$k_j$ and $p_j$ (normalized to killing norm $1$) where $k_j + i p_j$ are nontrivial roots. The
 remaining basis can be chosen as basis of ${\mathfrak k}_0$ and $\a$. We can organize the basis as the first $m$ vectors being $p_j$ followed by next $m$ elements as $k_j$ respectively. In these
basis $ad_a$ takes the block form
$$ \left [ \begin{array}{cc} A & 0 \\ 0 & 0  \end{array} \right ], $$ where
$$ A = \left [ \begin{array}{cc} 0 & \Lambda \\  -\Lambda & 0 \end{array} \right ],
\Lambda = \left [ \begin{array}{ccc} \lambda_1 & \hdots & \hdots \\ \vdots & \ddots & \vdots \\ 0 & 0 & \lambda_m \end{array} \right ] = i 2 \sigma_y \otimes \Lambda. $$

By performing a rotation by $S = \exp( -i \sigma_x \otimes I_m)$, we have
$$ S A S' = i 2 \sigma_z \otimes \Lambda = \left [ \begin{array}{cc} i \Lambda & 0 \\ 0 & -i \Lambda \end{array} \right ] $$
We define, $\tilde{S} = \left [ \begin{array}{cc}S & 0 \\ 0 & I_n  \end{array} \right ]$, The adjoint representation of $Ad_K(a)$, takes the form
$$ \Theta_1 \tilde{S}' \left [ \begin{array}{ccc} i \Lambda & 0 & 0\\ 0 & -i \Lambda & 0 \\  0 & 0 & 0 \end{array} \right ]\tilde{S} \Theta_1, $$ where $\Theta_1$ is matrix representation of $Ad_K(\cdot)$ over the chosen basis. It is orthonormal, as it preserves the killing norm. When $Ad_K$ is an automorphism of $\a$ , we have

$$ \tilde{S} ad_{Ad_K(a)} \tilde{S}' = \tilde{S} \Theta_1 \tilde{S}' \left [ \begin{array}{ccc} i \Lambda & 0 & 0 \\ 0 & -i \Lambda & 0 \\ 0 & 0 & 0 \end{array} \right ] \tilde{S} \Theta_1' \tilde{S}' = \left [ \begin{array}{ccc} i \tilde{\Lambda} & 0 & 0 \\ 0 & -i \tilde{\Lambda} & 0 \\ 0 & 0 & 0 \end{array} \right ] $$

Since eigenvalues are preserved by similarity transformation, there are only
finite possibilities $ \left [ \begin{array}{cc} i \tilde{\Lambda} & 0 \\ 0 & -i \tilde{\Lambda} \end{array} \right ] $, which means there are finite possibilities
for $ad_{Ad_K(a)}$ and $Ad_K(a)$. Hence Weyl points are finite and therefore number of
$Ad_K$ automorphisms of $\a$ are finite.

\begin{example}{\rm Let
$$ \g = -i \{ I_{\alpha}, S_\beta, I_{\alpha}S_{\beta} \} = \underbrace{-i {I_{\alpha}S_{\beta}}}_{\p} \oplus \underbrace{-i I_{\alpha}, S_{\beta}}_{\k},$$

$\a = -i \{I_\alpha S_\alpha \}$. Given $\alpha I_xS_x + \beta I_y S_y + \gamma I_z S_z$, the roots are
$-i \{ I_{y}S_{z} \pm I_z S_{y}  + i \frac{1}{2}(I_x \pm S_{x}) \}$, with value $\frac{\gamma \mp \beta}{2}$ , $-i \{ I_{z}S_{x} \pm I_x S_{z}  \mp i \frac{1}{2}(I_y \pm S_{y}) \} $ with value $\frac{\gamma \mp \alpha}{2}$ and $-i \{ I_{x}S_{y} \pm I_y S_{x}  + i \frac{1}{2} (I_z \pm S_{z})\} $, with value $\frac{\beta \mp \alpha}{2}$. Regular element is $|\alpha| \neq |\beta| \neq |\gamma|$. 
The fundamental roots are $\frac{\beta \pm \alpha}{2}$, $\frac{\gamma - \beta}{2}$.}
\end{example}

\begin{example}{\rm Let $$ \g = -i \{ I_{\alpha}, S_\beta, I_{\alpha}S_{\beta} \} = \underbrace{-i \{I_x, I_y, I_x S_\alpha, I_yS_{\beta}\}}_{\p} \oplus \underbrace{\{-i I_z, S_{\alpha}, I_{z}S_{\beta}\}}_{\k}, $$

$\a = -i \{I_x, I_xS_z \}$. Given $\alpha I_x  + \beta 2 I_x S_z = \frac{(\alpha+\beta)}{2} I_x(\frac{\Iden}{2} + S_z) + \frac{(\alpha -\beta)}{2} I_x(\frac{\Iden}{2} -S_z) $, the roots are

$-i \{ I_y(\frac{\Iden}{2} + S_z) + i I_z(\frac{\Iden}{2} + S_z) \}$, with value $\frac{\alpha + \beta}{2}$ ,
$-i \{ I_y(\frac{\Iden}{2} - S_z) + i I_z(\frac{\Iden}{2} - S_z) \}$ with value
$\frac{\alpha-\beta}{2}$,
$-i \{ 2I_xS_x + i S_y \}$ with value $\beta$,
$-i \{ 2I_yS_x + i 2I_z S_x \}$ with value, $\alpha$, $-i \{ 2I_yS_y + i 2I_z S_y \}$ with value $\alpha$
$-i \{ 2I_xS_y - i S_x \}$, with value $\beta$. Regular element is $|\alpha| \neq |\beta| \neq 0$.

The fundamental roots are $\beta$ and $\frac{\alpha - \beta}{2}$, with double root at $\alpha$ and $\beta$.}\end{example}


\begin{example}\label{eg:e-nroots}{\rm  Given the cartan decomposition $\g = \p \oplus \k$,  where, $\g = su(2n)$, $ \p = \left [ \begin{array}{cc} 0  & X \\ -X' & 0 \end{array} \right ],$  $\k = \left [ \begin{array}{cc} A  & 0 \\ 0 & B \end{array} \right ]$, where $Tr(A+B) = 0$, $A, B \in u(n)$, and $\a = \{ X = \Sigma \in diag(\lambda_i) \}$, $\lambda_i$ is real. We calculate, the roots $k + ip$.}  \end{example}

{\rm $A = A^1 + i A^2$, $B = B^1 + iB^2$ and $X = X^1 + iX^2$. Let $A^1_{ij}, B^1_{ij}$ be $1$, $-1$ in the $ij$ and $ji$ spot, $(i < j)$.
 Let $A^2_{ij}, B^2_{ij}$ be $1$, $1$ in the $ij$ and $ji$ spot, $(i < j)$. $X^{1}_{ij}, X^2_{ij}$ is $1$, in the $ij$ spot. $\Lambda^{\pm}_i$, is $1$ in $A^2_{ii}$ and $\pm 1$ in $B^2_{ii}$.

On $A^1_{ij}$, $B^1_{ij}$, $[\Sigma, .]$, takes the form

$$ \left [ \begin{array}{cc} -\lambda_j & \lambda_i \\ \lambda_i & -\lambda_j \end{array} \right ] $$ where
$X^1_{ij},
 X^1_{ji}$ are basis for the range. The eigenvalues and eigenvectors are
$\lambda_i - \lambda_j$ and $-(\lambda_i + \lambda_j)$ with eigenvectors
$[1, 1]'$ and $[1, -1]'$ respectively.

On $X^1_{ij}$, $X^1_{ji}$, $[\Sigma, .]$, takes the form
where
$$ \left [ \begin{array}{cc} \lambda_j & -\lambda_i \\ -\lambda_i & \lambda_j \end{array} \right ], $$ where
$A^1_{ij}, B^1_{ij}$ are basis for the range. The eigenvalues and eigenvectors are
$\lambda_j - \lambda_i$ and $(\lambda_i + \lambda_j)$ with eigenvectors
$[1, 1]'$ and $[1, -1]'$ respectively.

This gives $\frac{1}{\sqrt{2}} ( A^1_{ij} + B^1_{ij}) + i (X^1_{ij} + X^1_{ji})$ and  $\frac{1}{\sqrt{2}} ( A^1_{ij} - B^1_{ij}) + i (X^1_{ij} - X^1_{ji})$ are roots, with eigenvalues $-i(\lambda_i - \lambda_j)$ and $i (\lambda_i + \lambda_j)$, respectively.

On $A^2_{ij}$, $B^2_{ij}$, $[\Sigma, .]$, takes the form
$$ T_1 = \left [ \begin{array}{cc} -\lambda_j & \lambda_i \\ -\lambda_i & \lambda_j \end{array} \right ], $$ where
$X^2_{ij}, X^2_{ji}$ are basis for the range,

On $X^2_{ij}$, $X^2_{ji}$, $[\Sigma, .]$, takes the form
where
$$ T_2 = - T_1^T = \left [ \begin{array}{cc} \lambda_j & \lambda_i \\ -\lambda_i & -\lambda_j \end{array} \right ], $$ where $A^2_{ij}, B^2_{ij}$ are basis for the range.

Eigenvalues and eigenvectors of

$T_2 T_1 = - T_1^{T} T_1$, are $-(\lambda_i - \lambda_j)^2$, and $-(\lambda_1 + \lambda_j)^2$ with eigenvector
$[1, 1]$, $[1, -1]$, respectively, with
$T_1 [1, 1]' = (\lambda_i - \lambda_j) [1 -1]'$ and $T_1 [1, -1]' = -(\lambda_i + \lambda_j) [1, 1]'$.

This gives $\frac{1}{\sqrt{2}} ( A^2_{ij} + B^2_{ij}) + i (X^2_{ij} - X^2_{ji})$ and $\frac{1}{\sqrt{2}} ( A^2_{ij} - B^2_{ij}) + i (X^2_{ij} + X^2_{ji})$ are roots, with eigenvalues $-i(\lambda_i - \lambda_j)$ and $i (\lambda_i + \lambda_j)$, respectively.

We have a double root with eigenvalues $\lambda_i \pm \lambda_j$.

On $\Lambda^{+}$, we have $[\Sigma, \Lambda^{+}] = 0$ and $$[\Sigma, \Lambda^{-}_{k}]= -2 i \lambda_{k} X^2_{kk}, \ \ [\Sigma, i X^2_{kk}]= 2  \lambda_k \Lambda^{-}_{k}. $$

This gives $(\Lambda^{-}_k - X^2_{kk})$ are roots, with eigenvalues $i 2 \lambda_{k}$ respectively.}




\end{document}